\documentclass[11pt]{article}
\usepackage{graphicx,amsmath,amsfonts,changebar,color}
\usepackage{color}
\newcommand{\MP}[1]{\marginpar {\tiny {}}}

\newenvironment{Proof}{\noindent{\bf Proof:\ }}{
                        \vspace{0.25cm}\par}
\newenvironment{Prooflemma}{\noindent{\bf Proof:\ }}{
                        \vspace{0.25cm}\par}              
\newtheorem{proposition}{Proposition}[section]
\newtheorem{theorem}[proposition]{Theorem}

\newtheorem{lemma}[proposition]{Lemma}

\newtheorem{definition}[proposition]{Definition}
\newcommand{\0}{\vec{\mathbf 0}}
\newcommand{\1}{\vec{\mathbf 1}}
\newcommand{\8}{\vec{\mathbf v}}
\newcommand{\ren}{\stackrel{\sim}{\rightarrow}}

\newcommand{\Cons}{\mathrm{Cons}}
\newcommand{\AC}{\mathrm{AC}}
\newcommand{\N}{\mbox{\rm I\hspace{-.5em}N}}

\newcommand{\send}{\mbox{\em Send\/}}
\newcommand{\query}{\mbox{\em Query\/}}
\newcommand{\answ}{\mbox{\em Answer\/}}
\newcommand{\rec}{\mbox{\em Receive\/}}
\newcommand{\dec}{\mbox{\em Decide\/}}

\newcommand{\K}{{\,\leq_K\,}}
\newcommand{\C}{{\,\leq_C\,}}
\newcommand{\SC}{{\,\leq_{C^*}\,}}

\newcommand{\qedsymbol}{$\Box$}
\newcommand{\qed}{\nopagebreak \hfill \qedsymbol \nopagebreak}

\begin{document}

\title{Reductions in Distributed Computing\\ 
\vspace{0.3cm}
\Large{Part I: Consensus and Atomic Commitment Tasks} } 
\author{Bernadette Charron-Bost\thanks{Laboratoire LIX, 
	\'Ecole Polytechnique, 91128 	
	Palaiseau Cedex, France}}
\date{\empty} 
\maketitle

\begin{abstract}
 We introduce several notions of reduction in distributed computing,
and investigate reduction properties of two fundamental agreement tasks, 
namely Consensus and Atomic Commitment.

We first propose the notion of reduction ``{\em \`a la}  Karp'', an analog for  distributed 
	computing of the classical Karp reduction. 
We then define a weaker reduction
which is the analog of Cook reduction. 
These two reductions are called $K$-
reduction and $C$-reduction, respectively.
We also introduce the  notion of  $C^*$-reduction which has no counterpart in classical 
	(namely, non distributed) systems, and which naturally arises when 
	dealing with symmetric tasks.
	
We establish various reducibility and irreducibility theorems with respect to 
these three  reductions. Our main result is an incomparability statement for 
Consensus and Atomic Commitment tasks:  we show that they are incomparable 
with respect to the $C$-reduction, except when the resiliency degree is 1, in which 
case Atomic Commitment is strictly harder than Consensus. A side consequence 
of these results is that our notion of $C$-reduction is strictly weaker than the one 
of $K$-reduction,  even for unsolvable tasks.

\end{abstract}

\section{Introduction}

The purpose of this paper is  to develop a formalism for addressing the problems of 
	reduction in distributed computing, and to investigate reductions properties of 
	various agreement problems, namely,  Consensus and Atomic Commitment problems 
	(in Part~I), and their generalizations defined by the so-called $k$-Threshold Agreement
	problems~\cite{CL04} (in Part~II).

The notion of reduction plays a key role in the theory of computability.
A reduction from some problem $A$ to another one $B$ is a way of converting $A$ to $B$ in 
	such a way that a method for solving $B$ yields a method for solving $A$.
The existence of such a reduction establishes -- by definition -- that $B$ is at least as hard to
	solve as $A$,  or in other terms, that the {\em degree of 
	unsolvability}\footnote{Given some reduction relation, it 
	might seem natural to call an equivalence classes 
	-- with respect to the reduction -- which involve solvable problems a ``degree
	of solvability''; it is customary, however, to speak without exception 
	of ``degree of unsolvability''.}
	of $B$ is not less than the one of $A$.

	

Several notions of reducibility -- hence of degrees of unsolvability -- have been formally 
	defined and investigated in various frameworks.
Let us only mention the various kinds of effective reducibilities used in recursive functions theory
	(see for instance~\cite{Rog67}), and computation bounded reducibilities which play a key
	role in the theory of computational complexity, notably since the introduction of 
	polynomial-time bounded reducibilities by Cook~\cite{Coo71}
	and Karp~\cite{Kar72} (see also~\cite{LLS75} for a discussion and references on 
	polynomial-time reducibilities).
	
Concerning distributed computations, many reducibility results have been established and used to
	show that some problems are not solvable 
	(see for instance~\cite{FLM86, DDS87, MW87, CT96}).
Observe however that most of them are derived using an informal notion of reduction -- a significant
	exception being the work by Dwork and Skeen~\cite{DS84} on patterns 
	of communication.\footnote{The notion of reducibility 
	introduced in~\cite{DS84} is however much more restrictive than the one intuitively used 
	in the papers cited above: two problems $P_1$ and $P_2$ are equivalent with respect to the
	reducibility relation in~\cite{DS84} iff the sets of algorithms solving $P_1$ and $P_2$ 
	essentially  coincide, up to relabeling local states and padding messages.}
This lack of formal foundations for the reducibility notion in distributed computing is not a serious 
	issue in the proofs of the above reducibility results.
Indeed, they are established by means of constructive arguments which should be easily formalized
	in any sensible rigorous model for distributed computations and reducibility.
On the contrary, a formalized approach to reduction  in distributed computing is necessary for sound
	derivations of {\em irreducibility} results.

In the first sections of this paper, we develop such a formal approach.
As we are mainly interested in comparing the hardness of unsolvable tasks 
	-- this is similar to the study of polynomial-time reducibilities of problems which are not
	supposed to be solvable in polynomial time --  we need to refer,  in the definition of reducibility,
	to some {\it deus ex machina} for solving (algorithmically unsolvable) tasks. 
In other words, we grant the processes of a distributed system the ability to query a ``black box''
	 which supplies a correct  answer, magically solving  some specific distributed 
	coordination task.
Such a black box dedicated to solve some task is called, as in complexity theory,  an {\em oracle}. 
Our model for oracles takes into account  two specific 
	features of distributed computing:
	(i) an oracle has to synchronize and coordinate the queries from the different processes 
	in the system;
	(ii) in the context of systems where processes may exhibit failures, an oracle ought to answer
	even if some of the processes (the maximum number of which is the {\em resiliency degree}
	of the oracle) do not query it.
	
Relying on our formal definition of oracle, we may introduce {\em algorithms using oracles},  
	and define various notions
	of reduction between agreement tasks based on the latter.
Namely, we introduce  analogs for distributed
	systems of the many-one and Turing reductions in recursive functions theory.
Since their polynomial-time bounded versions in computational 
	complexity are the well-known 	 Karp and Cook reductions,  we call them 
	$K$-{\em reduction} and $C$-{\em reduction}.\footnote{In the 
	context of distributed systems, the
	terminology ``many-one'' and ``Turing'' reducibilities would be especially misleading.}
We also introduce notions of reducibility which have no counterpart in classical (namely, non 
	distributed) systems:  the $C^*$-{\em reduction} which arises  naturally when 
	dealing with symmetric tasks, and in Part~II, the 
	{\em Failure-Information reduction}, designed for 
	the study of failure resilient tasks.
	
Using this formalism, we may  derive rigorous reducibility and irreducibility theorems.
Our main result is an incomparability statement for Consensus and Atomic Commitment tasks: 
	we show that they are incomparable with respect to the $C$-reduction, except when the 
	resiliency degree is 1, in which case Atomic Commitment 
	is strictly harder than Consensus.
A side consequence of these results on  the comparison between  Consensus and Atomic Commitment 
	is that the notion of $C$-reduction is strictly weaker than the one of $K$-reduction, even for 	
	unsolvable tasks.
As shown in~\cite{LLS75}, a  similar situation arises for polynomial-time bounded 
	reducibilities between problems in {\bf P}.
Note however that comparing Karp and Cook reductions on {\bf NP} is still an open 
	problem.

Part~II of this paper will consider the class of $k$-{\em Threshold Agreement} tasks introduced in~\cite{CL04}
	which encompasses both Consensus and Atomic Commitment tasks.
We generalize the reducibility and irreducibility results established in Part~I to these new
	agreement tasks.
From these extensions, we derive new irreducibility results between
        Consensus tasks when varying the set of processes and the resiliency 
	degree.
	
Part~I is organized as follows.
In Section~2, following the general definition of decision tasks given in ~\cite{MW87}, 
	we define agreement tasks,  and then present the notion of symmetric agreement tasks.
In Section~3, we introduce our formal definition of oracle; technically, it is convenient to distinguish
	between an oracle and its name, which we call its {\em sanctuary} and which we need to
	introduce before the corresponding oracle.
We also explain the correspondence between our oracles and agreement tasks.
In Section~4, we follow the computational model developed in~\cite{CT96} to
	describe a computational model for message-passing systems in which processes may
	consult oracles.
Our model basically differs from the one in~\cite{CT96} by the fact that the computation unit
	called ``step'' is not  atomic any  more: taking a step in which it consults some oracle, a process 
	may be blocked after querying the oracle in the case the latter does not answer.
Section~5 defines the $K$-reduction, and establishes  $K$-reducibility and  irreducibility results.
In Section~6, we present the $C$-reduction, and its symmetrized version, called $C^*$-reduction.
In Section~7, we examine Consensus and Atomic Commitment tasks, and their reducibility relations
	when varying the number of processes in the system.
Our main results appear  in Section~8 in which we prove that  Consensus and Atomic 				
	Commitment tasks are generally incomparable. 
	
\section{Failure patterns and agreement tasks}

Our model of computation consists of a collection $\Pi$ of $n$ 	asynchronous processes, which communicate by exchanging messages.
Communications are point-to-point.
Every pair of processes is connected by a reliable channel.
We assume the existence of a discrete global clock to which processes 
	do not have access.
The range of the clock's ticks is the set of natural numbers, and is denoted
	by ${\cal T}$.

\subsection{Failures and failure patterns}\label{failures}

Processes may fail by crashing.
A {\em failure pattern} $F$ {\em for} $\Pi$ is a function 
	$F\, :\, {\cal T}\rightarrow 2^{\Pi}$, such that 
\begin{equation}\label{norecover}
\forall t\in {\cal T},\ F(t)\subseteq F(t+1).
\end{equation}
For any $t\in {\cal T}$, $F(t)$ represents the set of processes that have crashed 
	by time $t$.
If $p\notin F(t)$, we say that $p$ is {\em alive at time} $t$, and 
	condition~(\ref{norecover}) means that processes are assumed not to recover.
	
Process $p$ is {\em faulty} (with respect to $F$) if 
	$p\in Faulty(F)=\cup_{t\in {\cal T}} F(t)$;
	otherwise, $p$ is {\em correct} and $p\in Correct(F) = \Pi\setminus Faulty(F)$.

We only consider failure patterns with at least one survivor, that is the failure patterns $F$ 
	such that $|Faulty(F) < |\Pi|$.
The set of these failure patterns for $\Pi$ is denoted by ${\cal F}_{\Pi}$.

\subsection{Agreement problems and agreement tasks}\label{agpt}

We view an agreement problem as a mapping of possible inputs and failure patterns 
	to sets of allowable decision values.
Formally, let ${\cal V}$  be a set of input and output values, and $\Pi$ be a set
	of process names.
An {\em agreement problem} $P$ for $\Pi$ and ${\cal V}$ is given by a subset ${\cal V}_P$ 
	of ${\cal V}^{\Pi}$ and a mapping  
	$$P\, :\, {\cal F}_{\Pi}\times {\cal V}_P \longrightarrow 2^{\cal V}\setminus \{\emptyset\}.$$
Each element  $\vec{V}\in {\cal V}_P$ represents a possible initial assignment of input values 
	in ${\cal V}$ to the processes $p\in \Pi$ and is called an {\em input vector} of problem $P$.
For any $(F,\vec{V})\in {\cal F}_{\Pi}\times {\cal V}_P$, the non-empty subset $P(F,\vec{V})$ of 
	${\cal V}$ represents the set of allowable decision values with the input vector $\vec{V}$ and 
	the failure pattern $F$.

For any $v$ in ${\cal V}$, the constant mapping $\vec{V}$ defined by $\vec{V}(p)=v$, 
	for every $p\in \Pi$, is denoted by $\8$ (to simplify notation, we omit reference 
	to $\Pi$).

The simplest agreement problem for ${\Pi}$ and ${\cal V}$ is {\em Consensus}, 
	denoted $\Cons_{{\cal V},\Pi}$.
Its only requirement is that the decision value must be some process input value.
Formally, ${\cal V}_{\Cons_{{\cal V},\Pi}} = {\cal V}^{\Pi}$, and for each couple 
	$(F, \vec{V})\in{\cal F}_{\Pi}\times {\cal V}^{\Pi}$, the set 
	$\Cons_{{\cal V},\Pi}(F,\vec{V})$ of allowable decision values is defined as the set of elements of 		${\cal V}$ that occur in the input vector $\vec{V}$. 
	
In the case of the {\em binary consensus} problem for $\Pi$, simply denoted $\Cons_{\Pi}$, 
	we have ${\cal V}=\{0,1\}$, and the function $\Cons_{\Pi}$ is defined by:
\begin{itemize}
\item $\forall F\in {\cal F}_{\Pi}, \forall\, \vec{V}\in \{0,1\}^{\Pi} \setminus \{\0,\1\} :
			\Cons_{\Pi}(F,\vec{V})= \{ 0, 1 \};$
\item $\forall F\in {\cal F}_{\Pi} : \Cons_{\Pi}(F,\0) = \{ 0\} \mbox{ and } 
	\Cons_{\Pi}(F,\1) = \{ 1 \}.$
\end{itemize}

Another well-known agreement problem for $\Pi$ is {\em Atomic Commitment}, denoted $\AC_{\Pi}$.
It may be described as follows in terms of the previous definitions: ${\cal V}=\{0,1\}$, 
	${\cal V}_{AC_{\Pi}} = \{0,1\}^{\Pi}$, and for any 
	$(F,\vec{V})\in {\cal F}_{\Pi}\times \{0,1\}^{\Pi}$,
\begin{itemize}
\item $\AC_{\Pi}(F,\vec{V})= \{ 0 \} \mbox{ if } \vec{V}\neq \1,$
\item $\AC_{\Pi}(F,\vec{V})= \{ 1 \} \mbox{ if } \vec{V} = \1 \mbox{ and }
	Faulty(F)=\emptyset$,
\item $\AC_{\Pi}(F,\vec{V})= \{ 0,1 \} \mbox{ if } \vec{V} = \1 \mbox{ and }
	Faulty(F)\neq\emptyset$.
\end{itemize}
Classically, the input values are denoted by {\tt No} and {\tt Yes}, and 
	processes may decide on  {\tt Abort} or {\tt Commit}.
In the previous definition, we have identified {\tt Yes} and {\tt Commit} with 1, 
	and {\tt No} and {\tt Abort} with 0. 

For any set  $\Pi$ of $n$ process names, the data of an agreement problem $P$ for $\Pi$ 
	and of an integer $f$ such that $0\leq f \leq n-1$
	define an {\em agreement task}.
The  integer $f$ is called the {\it resiliency degree} of the task.
The tasks with the maximum resiliency degree are classically called {\em wait-free} tasks.

The distributed task defined  by the Atomic Commitment problem for $\Pi$
	and the resiliency degree $f$ will be denoted  $\AC(\Pi,f)$.
Similarly, we shall denote $\Cons(\Pi,f)$   the task defined by the Consensus problem and the 
	resiliency degree $f$.

\subsection{Renaming and symmetry}	
 
Let $\Pi$ and $\Pi'$ be two sets of $n$ process names, and let
	$\Phi : \Pi \ren \Pi'$ be a one-to-one mapping.
Such a map may be seen as a renaming of the processes in $\Pi$, and
	may be used to translate any input vector (or failure pattern, or agreement
	problem, or distributed task, ...) $X$ on the set of processes $\Pi$ to one 
	$^{\Phi} X$
	on the set of processes $\Pi'$.
These transformations under renaming are  bijective and satisfy the following 
	composition property: 
	if $\Pi''$ denotes a third set of $n$ processes, and $\Phi' : \Pi' \ren \Pi''$  
	is a one-to-one mapping, then 
	\begin{equation}\label{composename}
	^{\Phi'}(^{\Phi} X) = ^{\Phi' \circ \Phi} X.
	\end{equation}

Formally, these transformations are defined as follows:
\begin{itemize}
\item given any vector $\vec{V}$ in ${\cal V}^{\Pi}$,  the vector $^{\Phi}\vec{V}$
	 is the vector $\vec{V}\circ \Phi^{-1}$ in  ${\cal V}^{\Pi'}$; 
\item for any failure pattern $F$ for $\Pi$, we let  
	$^{\Phi} F: t\in {\cal T}\rightarrow \Phi(F(t));$
\item for any agreement problem $P$ for $\Pi$,  $^{\Phi}\! P$ is the agreement problem for 
	$\Pi'$ defined by:
	$$	{\cal V}_{^{\Phi}\!P} = \{ ^{\Phi}\vec{V}\ :\ \vec{V}\in {\cal V}_P\} \ \ \mbox{ and }\ \ 
		^{\Phi}\! P (^{\Phi}F, ^{\Phi}\vec{V}) = P(F,\vec{V}); $$
\item finally, for any agreement task $T=(P,f)$ for $\Pi$, we let $^{\Phi} T= (^{\Phi} P, f)$.
\end{itemize}

In the sequel, we denote by $\Cons(n,f)$ (resp. $\AC(n,f)$) the $f$-resilient task defined 
	by the Consensus (resp. the Atomic Commitment) problem for the set of process names
	$\Pi=\{1,\cdots, n\}$, that is:	
	$$\Cons(n,f)=\Cons \left  ( \{1,\cdots, n\},f \right )$$
and
	$$\AC(n,f)=\AC \left ( \{1,\cdots, n\},f \right ).$$
Clearly, for any set $\Pi$ of $n$ processes and for any renaming $\Phi : \{1,\cdots, n\} \ren \Pi$, 
	we have:
\begin{equation}\label{consn}
^{\Phi} \Cons(n,f) = \Cons (\Pi, f) 
\end{equation}
and
\begin{equation}\label{acn}
^{\Phi} \AC(n,f) = \AC (\Pi, f).
\end{equation}

Using transformations under renaming, we may formally define the {\em symmetry} of  an agreement 		problem $P$ or of a distributed task $T$ on some given set of processes $\Pi$.
Namely, $P$ (resp. $T$) is symmetric when, for any permutation $\sigma$ of $\Pi$, we have 
	$^{\sigma}\! P = P$ (resp., $^{\sigma} T = T$).
Clearly, $T=(P,f)$ is symmetric iff $P$ is.

In more explicit terms, the symmetry of $P$ means that, for any permutation $\sigma$ of $\Pi$, 
	${\cal V}_P$ is invariant by the permutation $\vec{V}\rightarrow ^{\sigma}\!\vec{V}$ of 
	${\cal V}^{\Pi}$,
	and for any failure pattern $F$ for $\Pi$
	and any input vector $\vec{V}$ in ${\cal V}_P$, we have 
	$P(^{\sigma}\!F, ^{\sigma}\!\vec{V})=P(F,\vec{V})$.
	
As a straightforward consequence of the composition property~(\ref{composename}), the symmetry 			property is invariant under renaming: if $P$ (resp. $T$) is  symmetric, then  for any renaming 
	$\Phi: \Pi\ren\Pi'$, $^{\Phi} P$ (resp.  $^{\Phi} T$) also is symmetric.
Observe finally that~(\ref{consn}) and (\ref{acn}) applied to permutations $\Phi$ of $\{1,\cdots,n\}$
	show that $\Cons(n,f)$ and $\AC(n,f)$ are symmetric.
By invariance of symmetry by renaming, this is equivalent to the symmetry of 
 	$\Cons(\Pi,f)$ and $\AC(\Pi,f)$ for any set $\Pi$ of processes. 
	
\section{Sanctuaries, oracles, consultations}\label{soc}
	
\subsection{Sanctuaries, consultations, and histories}\label{sanctuaires}

Informally, a distributed oracle for an agreement problem $P$ is a black box that 
	can be queried by processes with some input values for $P$, and that 
	is capable of reporting a solution to $P$ provided it has received  sufficiently 
	many queries.
Each oracle is identified by its name, which we call the {\em oracle's sanctuary}.

Formally, we fix a set of values ${\cal V}$, a set of process names $\Pi$, and a finite set 
	$\Sigma$ of sanctuaries.
Let $\Gamma : \Sigma\rightarrow 2^{\Pi}\setminus \{\emptyset\}$ be a function which
	assigns to each sanctuary $\sigma\in \Sigma$ a subset $\Gamma (\sigma)$ of $\Pi$
	which represents the set of processes allowed to consult $\sigma$.
The elements of  $\Gamma (\sigma)$ will be called the {\it consultants}  of $\sigma$.

An {\em event} at the sanctuary $\sigma\in \Sigma$ is defined as a tuple 
	$e=(\sigma,p,t,\tau,v)$,  where $p\in \Gamma (\sigma)$ is the {\em process 
	name of} $e$, $t\in {\cal T}$ is the {\em time} of $e$, 
	$\tau\in \{\mbox{Q}, \mbox{A}\}$ is the {\em type} of $e$, 
	and $v\in {\cal V}$ is the {\em argument value} of $e$.

Let $\sigma\in\Sigma$ be any sanctuary; a {\em history} $H$ of  $\sigma$ is a (finite or infinite) 
	sequence of events at $\sigma$ such that the times of events in $H$ form a non-decreasing list.
For any consultant $p$ of $\sigma$, the subsequence of all events in $H$ whose process names are 		$p$ will be denoted by $H|p$.
For any positive integer $k$, the $k${\em -th consultation} in $H$, denoted by $H_k$, is defined as
	the subsequence of all the events $e$ in $H$ such that for some $p\in \Gamma(\sigma)$,
	$e$ is the $k$-th query or the $k$-th answer event in $H|p$.
A history $H$ of the sanctuary $\sigma$ is {\em well-formed} if (i) for each process 
	$p\in \Gamma(\sigma)$, the first event in $H|p$, when $H|p$ is not empty, is  a query event, 
	(ii) each query event -- except possibly the last one -- is immediately followed by 
	an answer event, and (iii) each answer event -- except possibly the last one -- is 
	immediately followed by a query event. 
	
Let  $F$ be a failure pattern for $\Gamma(\sigma)$.
A history $H$ of the sanctuary $\sigma$ is {\em compatible with} $F$ if any process 
	that has crashed by some time does not consult $\sigma$ anymore;
	in other words, for any $(p,t)\in \Pi\times{\cal T}$ such that $p\in F(t)$, no event of the form 
	$(\sigma, p, t',-,-)$ with $t'\geq t$ occurs in $H|p$.\footnote{Throughout this paper, a ``-'' in a tuple 
	denotes an arbitrary value of the appropriate type.}
	
\subsection{Distributed oracles}\label{oracles}
	
For each sanctuary $\sigma\in \Sigma$, let ${\cal O}_{\sigma}$ be a function which maps 
	each failure pattern for $\Gamma(\sigma)$ to a set of well-formed histories of the
	sanctuary $\sigma$ which are compatible with $F$.
The function ${\cal O}_{\sigma}$ will be called the {\em oracle} of sanctuary $\sigma$.
Moreover, if  $P$ is an agreement problem for $\Gamma(\sigma)$, we shall say that 
	${\cal O}_{\sigma}$ {\em is an oracle suitable for} $P$ if for any failure pattern 
	$F$ for $\Gamma(\sigma)$, for any history $H\in {\cal O}_{\sigma}(F)$, and for any 
	positive integer $k$, the $k$-th consultation in $H$ satisfies the following two conditions:\\
	
\noindent{\em Agreement.} The oracle answers the same value to all processes.
Formally:
	$$(\sigma, -, - , \mbox{A}, d)\in H_k \wedge (\sigma, -, - , \mbox{A}, d')\in H_k 
	\Rightarrow d=d'.$$
	
\noindent{\em $P$-Validity.} If the oracle answers a value to some process, then this value 
	is allowed by $P$.
Formally, if $\vec{W}$ denotes the partial input vector defined by $H_k$ (namely
	$\vec{W}(p)=v$ iff $(\sigma,p,\mbox{Q},v)\in H_k$), and $\vec{V}$ any extension of 
	$\vec{W}$ in ${\cal V}_P$, any value $d$ answered by ${\cal O}_{\sigma}$ in $H_k$
	belongs to  $P(F, \vec{V})$.\\

Finally, we shall say that the oracle ${\cal O}_{\sigma}$ is $f$-{\em resilient} if for any failure pattern $F$ 
	for $\Gamma(\sigma)$, for any history $H\in {\cal O}_{\sigma}(F)$, and for any 
	consultation of ${\cal O}_{\sigma}$ in $H$ with at least $n-f$ query events, every 
	correct process finally gets an answer from ${\cal O}_{\sigma}$. 
Formally, ${\cal O}_{\sigma}$ is defined to be $f$-resilient if it satisfies:\\
	
\noindent $f${\em -Resilience.} For any failure pattern $F$ for $\Gamma(\sigma)$, 
	for any history $H\in {\cal O}_{\sigma}(F)$,
	and for any integer $k\geq 1$, the $k$-th consultation in $H$ satisfies:
$$\forall p\in Correct(F) \ :\  |\{e\in H_k : e=(\sigma, -, - , \mbox{Q}, -)\}|\geq |\Gamma(\sigma)|-f
	\Rightarrow (\sigma, p, - , \mbox{A}, -)\in H_k.$$

Observe that our oracles are suitable only for agreement problems.
However, it is straightforward to extend their definition to oracles suitable for 
	{\em decision problems}~\cite{BMZ90}.\footnote{In a decision problem,
	the sets of allowable decision values do not depend on failure patterns, and 
	processes may decide differently.
	{\em Renaming}~\cite{Att} and $k$-{\em Set Agreement problem}~\cite{Cha93} are two 
	well-known decision problems in which agreement is not required.}
	
\subsection{The oracle for an agreement task}\label{O.T}

Let $\Pi$ be a set of $n$ process names, and let $T$ be the task defined by some 
	agreement problem $P$ and some integer $f$, $0 \leq f \leq n-1$.
To these data, we may naturally attach some $f$-resilient oracle suitable for $P$, in the 
	following way.
Its sanctuary -- which, by definition, is a mere identifier -- will be $T$ itself, and it will be 
	suggestive and typographically convenient, to denote ${\cal O}.T$ instead of 
	${\cal O}_{T}$.
The set of consultants $\Gamma(\sigma)$ of the oracle ${\cal O}.T$ will be $\Pi$ itself, and for 
	any failure pattern $F$ for $\Pi$, we shall define ${\cal O}.T(F)$ as the set of all well-formed
	histories $H$ (of the sanctuary $T$) which are compatible with $F$, and satisfy the agreement,
	$P$-validity, and $f$-resilience conditions.
	
Clearly, ${\cal O}.T$ is the ``most general'' $f$-resilient oracle 
	for $P$, in the sense that for any $f$-resilient oracle ${\cal O}$ for $P$, 
	and for any failure pattern $F$, we have ${\cal O}(F)\subseteq {\cal O}.T(F)$.

The following properties of  the oracles for Consensus and Atomic Commitment tasks
	will be useful in the sequel.
They are straightforward consequences of the $\Cons$- and $\AC$-validity conditions
        (cf. Sections~\ref{agpt} and~\ref{oracles}).
\begin{enumerate}
\item[$\mathbf{O_{Cons}}$]  In any consultation of an oracle suitable for Consensus, 
	if all the queries have the same 
	value $v$, then the only possible answer of the oracle is $v$.
\item[$\mathbf{O_{AC}}$] In any consultation of an oracle suitable for Atomic Commitment, 
	the oracle is allowed  to answer 1 only if all processes query the oracle, and all the 
	query values are 1.
	
\end{enumerate}

\subsection{Related notions}\label{relnot}

The notion of oracle already appears at various places in the literature on distributed computing.
Indeed, it has been used in an informal way first for randomization~\cite{AT99b}  (see 
	also~\cite{CD89} where it occurs under the name of {\em coin}),  and then 
	for failure detectors~\cite{CT96}.
In both cases, an oracle is supposed to answer upon any query
	by some process.
Such an oracle has a maximal resiliency degree (namely $n-1$,  if $n$ is the number of processes
	which may query the oracle). 

This is not the only point in which random and failure detector oracles differ from ours.
In the case of  failure detector or randomization with private coins 
	-- as in Ben-Or's algorithm~\cite{Ben83}  -- the oracle is totally distributed and does not 
	coordinate the various queries from processes.
For this type of oracle, there is no notion of consultation.
Observe however that randomization with a global coin -- as in Bracha's algorithm~\cite{Bra87b} -- 		underlies a  notion of oracle which is closer to ours since all the processes see the same 
	outcome.

Interestingly, the fundamental concept of {\em shared object} introduced by Herlihy~\cite{Her91}
	has  some common flavor with our oracles.
Indeed, an object of type \texttt{consensus}~\cite{Jay97} in a system $\Pi$ with $n$ processes
	coincides with our oracle ${\cal O}.\Cons(\Pi,n-1)$. 
The generalization of the notion of shared object proposed by Malki {\em et al.} in~\cite{MMRT03}
	turns out to be yet closer: our $f$-resilient oracles for $\Pi$ actually correspond to  $f$-resilient 		shared objects of~\cite{MMRT03} with an {\em access list} $\Pi$ and only {\em one} operation.		

\section{Algorithms using oracles}
In this section, we fix ${\cal V}$, $\Sigma$, $\Pi$, two non-empty subsets $\Pi_1$ and $\Pi_2$, 
	of $\Pi$,  $\Gamma : \Sigma\rightarrow 2^{\Pi_2}\setminus \{\emptyset\}$, and a family 
	$({\cal O}_{\sigma})_{\sigma\in\Sigma}$ of oracles as defined in Section~\ref{soc}.
	
\subsection{Steps, events, and local histories}

We model the communication channels as a {\em message buffer}, denoted $\beta$,
	that represents the multiset of messages that have been sent but not yet delivered.
A message is defined as a couple $(p,m)$, where $p$ is the name of the destination process, 
	and $m$ is a message value from a fixed universe $M$.
	
An {\em algorithm for} $\Pi_1$ {\em using the oracles of the sanctuaries in} $\Sigma$
	is a function $A$ that maps  each process name $p\in \Pi_1$ to a 
	deterministic automata $A(p)$. 
The computation locally proceeds in {\em steps}.
Each step of $A(p)$ consists in a series of different phases:\\

\noindent{\em Message Receipt.} Process $p$ receives a single message of the form $(p,m)$ from $\beta$. \\

\noindent{\em Oracle Query.} Process $p$ queries a single oracle ${\cal O}_{\sigma}$, 
	$p \in \Gamma(\sigma)$, with some value $v\in {\cal V}$.\\
	 
\noindent{\em Oracle Answer.} Process $p$ gets an answer $d\in {\cal V}$ from the oracle that 
	$p$ consults in this step.\\
	
\noindent{\em State Change.} Process $p$ changes its local state, and sends a message 
	to a single process or sends no message, according to the automaton $A(p)$.
	These actions are  based on $p$'s state at 
	the beginning of the step, the possible message received in the step, and the possible 
	value answered by the oracle.\\

In every step, $p$ may skip the two intermediate phases ($p$ consults 
	no oracle); it may also skip the first phase ($p$ receives no message).
So there are four kinds of steps with one, two, three, or four phases,
	whether $p$ receives or not a message, and whether it consults or
	not an oracle.

The message actually received by $p$ in the Message Receipt phase is chosen
	nondeterministically amongst the messages in $\beta$ that are addressed 
	to $p$.
Process $p$ may receive no message even if $\beta$ contains messages that are 
	addressed to $p$.
Indeed,  we model asynchronous systems, where messages 
	may experience arbitrary (but finite) delays.
	
Besides, the fact that $p$ is allowed or not to consult an oracle in some step 
	is totally determined by the local state of $p$ at the beginning of the step.
Moreover, in the case of a local state in which $p$ consults an oracle, the
	name of the oracle (i.e., the sanctuary) is also completely determined by the local 
	state.
A step is thus uniquely determined by (1) the name $p$ of the process that takes
	the step, (2) the message $m$ (if any) received by $p$ during that step, and in the case $p$ 
	consults an oracle in the step, (3) the value answered by the oracle.
We may therefore identify a step with a triple $[p,m,d]$, where 
	$p\in \Pi_1$, $m\in M\cup\{\mbox{null}\}$ with $m=\mbox{null}$ if $p$ receives no message
	in the step, and $d\in {\cal V}\cup\{\bot\}$ with 
	$d=\bot$ if $p$ consults no oracle in the step.
Given a local state $state_p$ of $p$, we say that the step $s=[p,m,d]$ {\em is feasible 
	in} $state_p$ in the two following cases:
	\begin{enumerate}
	\item $d$ is in ${\cal V}$ and  $p$ has to consult an oracle in $state_p$;
	\item  $d=\bot$ and $p$ is not allowed to consult any oracle in 
	$state_p$.
	\end{enumerate}
We denote by $s(state_p)$ the unique state of $p$ that results when $p$ performs 
	the step $s$ in the state $state_p$.

This description of a step leads to generalize the definition of events given in 
	Section~\ref{oracles}, and to consider two new types of events:
	$(\beta,p,t,\mbox{R},m)$
	and $(\beta,p,t,\mbox{S},m')$ -- $R$ stands for ``Receive'', and $S$ for ``State change'' --
	where  $p\in\Pi_1$, $t \in {\cal T}$, $m\in M$, and $m'\in M\cup \{null\}$.
A step is thus a series of one, two, three, or four events of the following 
        form:
	\begin{enumerate}
	\item $\langle (\beta,p,t,\mbox{S},m')\rangle$
	\item $\langle (\beta,p,t,\mbox{R},m);(\beta,p,t,\mbox{S},m')\rangle$
	\item $\langle (\sigma,p,t,\mbox{Q},v);(\sigma,p,t,\mbox{A},d);
	  (\beta,p,t,\mbox{S},m')\rangle$ 
	\item $\langle (\beta,p,t,\mbox{R},m);(\sigma,p,t,\mbox{Q},v);
	(\sigma,p,t,\mbox{A},d);(\beta,p,t,\mbox{S},m')\rangle$.
	\end{enumerate}

\subsection{Histories and runs}
	
A {\em history of process} $p$ is a (finite or infinite) sequence of events
	whose process names are $p$, and such that the times of events in this sequence
	form a non-decreasing list.
A history $H_p$ of process $p$ is  {\em well-formed} if the events in $H_p$ can be 
	grouped to form a sequence of steps, except possibly the last events which may 
	only form a prefix of a step with the message receipt and oracle query phases
	(the oracle answer and state change phases may be both missing).
The resulting sequence of {\em complete} steps in $H_p$ is denoted $\overline{H_p}$.

For every failure pattern $F$ for $\Pi_1$ and every sanctuary $\sigma\in\Sigma$, 
	we define the failure pattern $F_{\sigma}$ for 
	$\Gamma(\sigma)$ by
	$$F_{\sigma}(t)=\left ( F(t)\cap \Gamma(\sigma)\right ) \cup 
	\left(\Gamma(\sigma)\setminus \Pi_1\right),$$
	i.e., $F_{\sigma}$ consists of the consultants of $\sigma$ which are either faulty 
	with respect to $F$  or not in the membership of $\Pi_1$.

Let  $F$ be a failure pattern for $\Pi_1$;  a  history $H_p$ of process $p$ is said to be 
	{\em compatible with} $F$ if any process that has crashed by some time  in $F$ performs 
	no step afterwards; in other words, for any $(p,t) \in \Pi_1\times {\cal T}$ such that $p\in F(t)$, 
	no event of the form $(-, p, t',-,-)$ with $t'\geq t$ occurs in $H_p$.
	
A {\em history} $H=(e_i)_{i\geq 1}$ {\em of the algorithm} $A$ is a (finite or infinite) 
	sequence of events such that their times  $(t_i)_{i\geq 1}$ form a non-decreasing 
	sequence in ${\cal T}$. 
The subsequence of all events in $H$ whose process name is $p$ is denoted by $H|p$.
Similarly, $H|\sigma$ denotes the subsequence of events in $H$ related to the sanctuary 
	$\sigma$.
	
We assume that initially, the message buffer $\beta$ is empty and
	every process $p$ is in an initial state of $A(p)$.
	
From history $H$, we inductively construct the sequence 
	$(state_{\beta}[i])_{i\geq 0}$ in the following way:
	(a) $state_{\beta}[0]=\emptyset$, and (b) if $e_i=(\beta,-,t_i,\mbox{R},m)$, 
	then $state_{\beta}[i]= state_{\beta}[i-1]\setminus \{ m\}$, and 
	if  $e_i=(\beta,-,t_i,\mbox{S},m')$ with $m'\neq null$, then 
	$state_{\beta}[i]= state_{\beta}[i-1]\cup\{ m'\}$;
	otherwise, $e_i$ is an event that does not modify $state_{\beta}[i]$, 
	i.e., $state_{\beta}[i]= state_{\beta}[i-1]$. 
	
A {\em run} of $A$ is a triple $\rho=<\!\! F, I, H \!\!>$ where $F$ is a failure pattern 
	for $\Pi_1$, $I$ is a function mapping each process $p$ to an initial state 
	of $A(p)$, and $H$ is a history of $A$ that satisfy the 
	following  properties R1--6:
\begin{description}
\item[R1] For every sanctuary $\sigma\in \Sigma$, the subhistory 
	$H|{\sigma}$ is a  history of the sanctuary $\sigma$ which is both well-formed
	and compatible with $F_{\sigma}$.
	Formally, 
	$$\forall\sigma\in\Sigma : H|{\sigma}\in {\cal O}_{\sigma}(F_{\sigma}).$$
\item[R2] For every process $p\in \Pi_1$, the subhistory 
	$H|p$ is a history of the process $p$ which is both well-formed and compatible with $F$.
\item[R3] Every message that is delivered by $p$ has been previously sent to $p$.
Formally, 
	$$\forall m\in M :  (\beta,p,t_i,\mbox{R},m)\in H\Rightarrow
	(m=(p,-)\wedge m\in state_{\beta}[i-1]).$$
\item[R4] Every step in $H$ is feasible. 
Formally, $state_p[0]=I(p)$ and for every process $p$, $\overline{H|p}[1]$ is feasible 
	in $state_p[0]$, $\overline{H|p}[2]$ is feasible in 
	$state_p[1]= \overline{H|p}[1](state_p[0])$, etc ...
\end{description}

To state our two last conditions, we need to introduce the notion of ``process 
	locked in a sanctuary''.
	
An answer event {\em matches} a query event if their process names and their 
	oracle names (sanctuaries) agree.
A query event is {\em pending} in a history if no matching answer event follows 
	the query event.
We say that process $p$ {\em is locked} in the sanctuary  $\sigma$ during 
	$\rho=<\!\! F, I, H \!\!>$ if $p\in Correct(F)$ and there is a pending 
	query event of the type $(\sigma, p, -, \mbox{Q}, -)$ in $H$.
We denote by $Locked(\rho)$ the set of processes in $\Pi_1$ which are locked
	in some sanctuary of $\Sigma$ during $\rho$.
	
\begin{description}
\item[R5] Every correct process that is not locked in any sanctuary takes an 
        infinite number of steps.
Formally,
$$\forall p\in \Pi_1, \forall i : p\in Correct(F)\setminus Locked(\rho)
	\Rightarrow \exists j>i: (-,p,t_j,-,-)\in H.$$
\item[R6] Every message sent to a correct process that is locked in no sanctuary is 
	eventually received. 
Formally, 
$$\forall p\in \Pi_1, \forall i : 
(p\in Correct(F)\setminus Locked(\rho)\ \wedge\ m=(p,-)\in state_{\beta}[i])\Rightarrow $$
$$(\exists j>i: (\beta,p,t_j,\mbox{R},m)\in H).$$

\end{description}

Observe  that conditions R1--6 are not independent (for instance, the compatibility
	requirement in R1 is implied by R2).
	
\subsection{Terminating algorithms}\label{term}

So far, we have not made any provision for {\em process stopping}.
It is easy, however, to distinguish some of the process states  as {\em halting states},
	and specify that no further activity can occur from these states.
That is no messages are sent and the only transition is a self-loop.\footnote{Note that these 
	halting states do not play the same role as they do  in classical finite-state automata theory.
	There, they generally serve as {\em accepting states}, which are used to determine which
	strings are in the language computed by the machine.
	Here, they just serve to model that processes halt.}
An algorithm $A$ is said to be {\em terminating in the presence of $f$ failures}  if  in any run of $A$
	with at most $f$ failures, every correct process eventually reaches a halting state.

It is important to notice the difference between the fact that a process may make a decision and the 
	fact that it may cease participating to the algorithm, that is it may halt~\cite{DS84}.
Indeed, as shown by Taubenfeld, Katz, and Moran~\cite{TKM89} for initial crashes or by Chor and 
	Moscovici~\cite{CM89} for the randomized model, solvability results for decision tasks highly
	depend on whether processes are required to terminate (after making a decision) or not.
	
The definition of solvability given above does not include the termination requirement.
However, in the case of agreement tasks, solvability does imply solvability with termination.
To show that, it suffices to see that any algorithm solving an agreement task $T$ can be 
	translated into a terminating algorithm which solves $T$ too.
Let $A$ be any algorithm solving $T$; we transform $A$ in  following way:
\begin{enumerate}
\item as soon as a process makes a decision in $A$, it sends its decision value to all
	and then it halts;
\item upon the receipt of a decision notification, a process stops running $A$, decides
	on the value it has received.
In turn, it sends its  decision value to all, and then halts.
\end{enumerate}
Clearly, the resulting algorithm $B$ solves $T$ and every correct process eventually  terminates.
The corresponding definition of the automata $B(p)$ from $A(p)$ is trivial, and so omitted.

\subsection{Algorithms for agreement tasks}\label{algora}

In the context of  agreement problems, each process $p$ has an initial value 
	in ${\cal V}$ and must reach an irrevocable decision on a value in ${\cal V}$.
Thus for an agreement problem, the algorithm of process $p$, $A(p)$,  has 
	distinct initial states $s_p^v$ indexed by $v\in {\cal V}$, $s_p^v$ signifying that 
	$p$'s initial value is $v$. 
The local algorithm $A(p)$ also has disjoint sets of decision states $S_p^d$,
	$d\in {\cal V}$.

We say that algorithm $A$ for $\Pi_1$ using oracles of the $\Sigma$ sanctuaries 
	{\em solves the agreement task} $(P,f)$ for $\Pi_1$ if every run $\rho=<\!\! F, I, H \!\!>$
	of $A$ where $F$ is a failure pattern with at most $f$ failures satisfies:\\

\noindent{\em Termination.} Every correct process eventually decides some value.
Formally,
$$\forall p\in Correct(F), \exists i: state_p[i]\in \bigcup_{d\in {\cal V}} S_p^d.$$

\noindent{\em Irrevocability.} Once a process makes a decision, it remains decided on
	that value.
Formally, 
$$\forall p\in \Pi_1, \forall d\in {\cal V}, \forall i,j :  
	(i\leq j \ \wedge \ state_p[i]\in S_p^d)\Rightarrow state_p[j]\in S_p^d.$$
	
\noindent{\em Agreement.} No two processes decide differently.
Formally,
$$\forall p, p'\in \Pi_1,\forall i, \forall d,d'\in {\cal V} : 
	(state_p\in S_p^d \wedge state_{p'}\in S_{p'}^{d'})\Rightarrow d=d'.$$
	
\noindent $P$-{\em Validity.} If a process decides $d$, then $d$ is allowed by $P$.
Formally, let $\vec{V}$ denotes the vector of initial values defined  by $I$.
$$\forall p\in \Pi_1, \forall d\in {\cal V}: 
	(\exists i:  state_p[i]\in S_p^d)\Rightarrow d\in P(F,\vec{V}).$$

\section{Reduction \`a la Karp: ${\mathbf K}$-reduction}\label{K-red}

We need a precise definition of what it means for a task to be {\em at least as hard
	as} another one.
For that, we first propose the notion of reduction {\em \`a la} Karp, an
	analog for distributed computing of the classical Karp reduction.
Informally, task $T_1$ {\em $K$-reduces} to task $T_2$ if, to solve $T_1$, we just 
	have to transform the input values for $T_1$ into a set of inputs for $T_2$, 
	and solve $T_2$ on them.
When this holds, we shall say that $T_2$ is (at least) as hard as $T_1$.
We shall prove that in synchronous systems, Consensus
	tasks are strictly harder than Atomic Commitment tasks, with respect to $K$-reduction.

\subsection{${\mathbf K}$-reduction}

\begin{definition}\label{Kred}
Let  $T_1$ and $T_2$ be two tasks for a set $\Pi$ of processes. 
We say that $T_1$ is $K$-{\em reducible} to $T_2$, and we note $T_1\K T_2$, if
	there is an algorithm for $T_1$ in which each correct process $p$ in $\Pi$ (1) transforms 
	any input value $v_p$ into some value $w_p$ without using any oracle, and (2) 
	queries the oracle ${\cal O}.T_2$ with  $w_p$, gets an answer value $d$ from the oracle, 
	and finally decides on $d$. 
The first part $R$ of the algorithm which transforms every vector $\vec{V}$ for
        $T_1$ into 
	the partial vector $\vec{W}=R(\vec{V})$ using no oracle is a terminating algorithm called a 
	$K$-{\em reduction from} $T_1$ to $T_2$.
\end{definition}

As explained in the Introduction, we expect that if some task is reducible to a second solvable task, 
	then we can obtain a solution for the first one.
This is satisfied by $K$-reduction, as shown by the following proposition.

\begin{proposition}
If $T_1$ $K$-reduces to $T_2$ and $T_2$ is a solvable task, then $T_1$ is solvable.
\end{proposition}

\begin{Proof}
If $T_2$ is solvable, then there is an algorithm $A$ using no oracle which solves $T_2$.
Thus we may replace the oracle ${\cal O}.T_2$ by $A$, just after $R$ terminates.
The resulting algorithm uses no oracle and solves $T_1$.\qed
\end{Proof}
	
Clearly $K$-reduction is reflexive; 
	moreover it is transitive, namely if $T_1$, $T_2$, and $T_3$ are three tasks 
	for $\Pi$ such that 
	$T_1\K T_2$ and  $T_2\K T_3$, then  $T_1\K T_3$.
Thus it orders tasks with respect to their difficulty.

Let  $T_1$ and $T_2$ be two agreement tasks, and let 
        $f_1$ and $f_2$ denote their resiliency degrees, respectively.
From Definition~\ref{Kred}, it follows that if $T_1\K T_2$, then 
        ${\cal O}.T_2$ definitely answers when $f_1$ processes do not query it.
This implies that $f_1\leq f_2$.

Conversely, it is immediate to see that the more a task is resilient, the harder it is to solve it.
Formally, if  $T_1$ and $T_2$ are defined by the {\em same} agreement 
         problem and $f_1\leq f_2$, then $T_1\K T_2$.
Therefore, in the case of two agreement tasks defined by the same problem, 
         task  $T_1$ $K$-reduces to $T_2$ if and only if $f_1\leq f_2$.
	
Notice that the key point for proving some $K$-reduction between two agreement tasks lies
	in the validity condition.
Indeed, let $T_1=(P_1,f_1)$ and $T_2= (P_2,f_2)$ be two agreement tasks for some set $\Pi$
	of processes.
Assume that $f_1\leq f_2$ (cf. discussion above).
Let $R$ be any algorithm running on $\Pi$ in which, starting from any input vector $\vec{V}$
	for $T_1$, every correct process $p$ eventually outputs some value $w_p\in {\cal V}$.
We consider the algorithm resulting from the query of the oracle for $T_2$ with the output 
	values of $R$, and study whether this algorithm solves or not $T_1$.
Irrevocability is obvious; agreement is also trivial ($T_1$ and $T_2$ share this condition).
Because no process may be blocked in $R$ and because ${\cal O}.T_2$ is an $f_2$-resilient oracle,  
	termination is  guaranteed in any run with at most $f_2$, and so $f_1$, failures.
Hence, showing $R$ is a $K$-reduction from $T_1$ to $T_2$ actually consists in proving that
	the answer given by ${\cal O}.T_2$ ensures that the $P_1$-validity condition
	is satisfied.
	


%
%

\subsection{K-reducibility between Consensus and Atomic Commitment tasks}

We first establish a $K$-reduction result in the particular case of synchronous systems.
Recall that in such systems, one can emulate a computational model in which computations 
	are organized in {\em rounds} of information exchanges.
On each process, a round consists of message sending to all processes, 
	receipt of all the messages sent to this process at this round,
	and local processing (see Chapter 2 in~\cite{Lyn96} for a detailed presentation 
	of this computational model).
	
\begin{theorem}\label{synKred}
In the synchronous model, for every integers $n, f$ such that $0 \leq f\leq n-1$, $\AC(n,f)$ 
	is $K$-reducible to $\Cons(n,f)$.
\end{theorem}

\normalsize

\begin{figure*}[t]\small
$\hrulefill$
\begin{tabbing}
mm\=mm\=mm\=mm\=mm\=mm\=mm\=mm\=mm\=mm\=mm\= \kill

$\mathbf{Code\ for\ process\ p:}$\\ \\
\> send $\langle v_p \rangle$ to all\\
\> Receive all messages $\langle v_q \rangle$ sent to $p$\\
\> {\bf if} {received $n$ messages with value 1} {\bf then} $w_p:=1$\\
\> {\bf else} $w_p:=0$ 
\end{tabbing}
\vspace{-0.5cm}
\caption{A $K$-reduction from $\AC(n,f)$ to $\Cons(n,f)$ in the synchronous model.}
\label{synalgored}
$\hrulefill$
\end{figure*}

\begin{Proof}
Consider the one round algorithm $R$ in Figure~\ref{synalgored} which transforms 
	every input value $v_p$ into $1$ if process $p$ detects no failure
	(i.e., $p$ receives exactly $n$ messages) and all the values that $p$ 
	receives are equal to 1; otherwise, $R$ transforms $v_p$ into 0.
We claim that $R$ is a $K$-reduction from $\AC(n,f)$ to $\Cons(n,f)$.

As mentioned above, we just have to address validity.
If no failure occurs, then every process receives $n$ messages in the one round 
	algorithm $R$.
Therefore if all the $v_p$'s are equal to 1 and no process fails, 
	then all the $w_p$'s are set to 1.
Every process that is still alive queries the oracle for $\Cons(n,f)$ with value 1,
	and so by the validity condition of consensus, the oracle answers 1.
On the other hand, suppose at least one process starts with 0.
Each process $p$ receives less than $n$ messages in $R$ or receives at least one message 
	with value 0.
In both cases, $w_p$ is set to 0.
All processes query the $\Cons(n,f)$ oracle with value 0; by the validity condition of
	consensus, the oracle definitely answers 0. 
Therefore the validity condition of Atomic Commitment is satisfied.\qed
\end{Proof}

Conversely, we prove that if $f\geq 1$ then $\Cons(n,f)$ is not $K$-reducible to $\AC(n,f)$, and so
	$\Cons(n,f)$ is strictly harder to solve than $\AC(n,f)$ in synchronous systems.

\begin{theorem}\label{nok-TAgred}
For any integers $n, f$ such that $1\leq f \leq n-1$, $\Cons(n,f)$ is never $K$-reducible to $\AC(n,f)$,
	even in synchronous systems. 
\end{theorem}

\begin{Proof}
For the sake of contradiction, suppose that there exists a $K$-reduction $R$
	from $\Cons(n,f)$ to $\AC(n,f)$ in the synchronous model, and consider the 
	resulting algorithm for $\Cons(n,f)$.
	
We consider a failure free run $\rho$ of this algorithm which starts with the input vector $\1$;
	let $p$ be the name of the process which terminates $R$ last in this run,  
	and let $r_p$ denote the round number when $p$ completes 
	the computation of $w_p$ in this run.
Let $F$ denote the failure pattern such that all processes are correct, except 
	$p$ which crashes just at the end of round $r_p$.
Now there is a run $\rho'$ of the algorithm for $\Cons(n,f)$ whose failure pattern is $F$, which 
	starts with the input vector $\1$, and such that every process has the same behavior 
	by the end of round $r_p$ in $\rho'$ as in $\rho$.
In $\rho'$, process $p$ does not query the $\AC(n,f)$ oracle which therefore definitely 
	answers 0 (cf.  property $\mathrm{O_{AC}}$ in Section~\ref{O.T}).
Consequently, the validity condition of consensus is violated in $\rho'$, 
	a contradiction when $f\geq 1$. 
	
This proves that $\Cons(n,f)$ is not $K$-reducible to $\AC(n,f)$ in synchronous systems, and so 
	in asynchronous systems. \qed
\end{Proof}

\section{Reductions \`a la Cook: C-reduction and ${\mathbf C^*}$-reduction}

In Section~\ref{K-red}, we introduced the notion of $K$-reducibility as a specific
	way of using a solution to one task to solve other tasks: if $T_1$ is 
	$K$-reducible to $T_2$, and we have a solution for $T_2$, we obtain a 
	solution for $T_1$ just by transforming the input values for $T_1$ into 
	input values for $T_2$.
Such a reducibility notion is very restrictive: just {\em one} solution for
	$T_2$ can be used to design a solution for $T_1$, and just {\em in the end}. 
	
We now propose some weaker notion of reduction in which every process is allowed 
	to query the oracle several times, and not just in the end as with $K$-reduction. 
Actually, we define two such notions of reduction: a first one which applies to arbitrary
	tasks for some given sets of processes, and a second one which makes sense for
	symmetric tasks.
These are analogs  for distributed computing of the classical Cook reduction, and 
	will be called $C$- and $C^*$-{\em reduction}.

\subsection{${\mathbf C}$-reduction}

Consider the following data:
\begin{itemize}
\item a finite set of process names $\Pi$;
\item a family $\{\Pi_2^{\sigma}\ :\ \sigma\in \Sigma\}$ of subsets of $\Pi$, indexed by 
	a finite set $\Sigma$ (the sanctuaries), and, for any $\sigma\in\Sigma$, a task 
	$T_2^{\sigma}$ for $\Pi_2^{\sigma}$;
\item a finite subset $\Pi_1$ of $\Pi$, and a task $T_1$ for $\Pi_1$.
\end{itemize}

\begin{definition}\label{Cred}
We say that $T_1$ is $C$-{\em reducible} to $\{T_2^{\sigma}: \sigma\in \Sigma\}$,
	and we note 
	$$T_1 \C \{T_2^{\sigma}: \sigma\in \Sigma\} $$
	if there is an algorithm $R$ for $T_1$ using the oracles 
	$\{{\cal O}.T_2^{\sigma} : \sigma\in \Sigma\}$.
The algorithm $R$ is called a $C$-{\em reduction from} $T_1$ to 
	$\{T_2^{\sigma}: \sigma\in \Sigma\}$.
\end{definition}

Often, we deal with a set of sanctuaries reduced to a singleton:
	the family $\{T_2^{\sigma}: \sigma\in \Sigma\}$ is then given by one task $T_2$,
	and we simply say that ``$T_1$ is $C$-reducible to $T_2$'', and write 
	$T_1\C T_2$.

This notion of $C$-reducibility is transitive in the following strong sense: 
Consider a set of sanctuaries $T$, and for any $\tau\in T$, a set of ``affiliated sanctuaries''
	$\Sigma(\tau)$.
Let $$\Sigma^T = \bigcup_{\tau\in T} \{\tau\}\times \Sigma(\tau).$$
Assume that a task $T_1$ is $C$-reducible to a family of tasks $\{T_2^{\tau} : \tau\in T\}$,
	and that, for any $\tau\in T$, the task $T_2^{\tau}$ is $C$-reducible to a family of 
	tasks $\{T_3^{\sigma,\tau} : \sigma\in\Sigma(\tau)\}$.
Then $T_1$ is $C$-reducible to the family of tasks 
	$\{T_3^{\sigma,\tau} : (\tau,\sigma)\in\Sigma^T\}$.

In particular, restricted to single tasks, $C$-reducibility is transitive in the usual sense:
	if $T_1\C T_2$ and $T_2\C T_3$, then $T_1\C T_3$.
It is also clearly reflexive.
	
$C$-reduction satisfies our intuitive concept of reducibility as shown by the following proposition.

\begin{proposition}\label{Csolv}
If $T_1$ $C$-reduces to $\{T_2^{\sigma}: \sigma\in \Sigma\}$ and every task $T_2^{\sigma}$ is a 			solvable task, then $T_1$ is solvable.
\end{proposition}
\begin{Proof}
Let $R$ be a $C$-reduction from  $T_1$  to $\{T_2^{\sigma}: \sigma\in \Sigma\}$.
Since  $T_2^{\sigma}$ is solvable, there exists an algorithm $B^{\sigma}$ using no oracle which solves 
	$T_2^{\sigma}$.
As explained in Section~\ref{term}, we may suppose that  $B^{\sigma}$ is a terminating algorithm.

Let  $p$ be a process in $\Pi_1$, and let $state_p$ be any state of  $R(p)$  in which
	$p$ consults some oracle ${\cal O}.T_2^{\sigma}$  with the query value $v$.
Consider a transition of $R(p)$ from $state_p$  corresponding to step $[p,m,d]$.
Since $B^{\sigma}$ terminates,  there is no problem to replace  this transition in $R(p)$ 
	by the (possibly empty) sub-automata of $B^{\sigma}(p)$ consisting of the states which are 			reachable from the initial state $s^v_p$  and leading to some halting states in $S^d_p$.
Thus we may replace the algorithm for $T_1$ using the set of oracles 
	$\{{\cal O}.T_2^{\sigma} : \sigma\in \Sigma\}$
	by an ordinary algorithm
	(using no oracle) that solves $T_1$.\qed
\end{Proof}

Combined with the  transitivity of $C$-reduction, this latter proposition shows  that like $K$-reducibility, 		$C$-reducibility  orders tasks with respect  to their difficulty. 
	
From Definition~\ref{Cred}, it is straightforward that $K$-reducibility is at least as strong 
	as $C$-reducibility: if $T_1$ and $T_2$ are two tasks for $\Pi$ such 
	that $T_1 \K T_2$, then  $T_1 \C T_2$.
However, some major differences between $K$- and $C$-reducibility should be emphasized.
Firstly, the flexible use of oracles in the definition of $C$-reduction allows us to 
	compare tasks for different sets of processes, whereas any two tasks comparable with 
	respect to the $K$-reduction are necessarily tasks for the same set of processes.
Secondly, note that if $T_1$ is any solvable task, then it is $C$-reducible to any task $T_2$;
	this would not hold for $K$-reducibility (cf. Theorem~\ref{nok-TAgred}).
	
This latter remark shows that for any two solvable tasks $T_1$ and $T_2$, we have
	both $T_1\C T_2$ and  $T_2\C T_1$.
In other words, two solvable tasks are equivalent with respect to $C$-reducibility.
Actually, $C$-reduction discriminates unsolvable tasks and is aimed to determine 
	{\em unsolvability degrees}.
Since we focus on agreement problems which are all solvable in the absence of 
	failure, from now on we shall assume that the resiliency degree of tasks is at least~1.

\subsection{${\mathbf C^*\!}$-reduction}

Let $\Pi_1$ and $\Pi_2$ be two sets of $n_1$ and $n_2$ processes, respectively. 
Let $T_1$ be a task for $\Pi_1$, and $T_2$ be a symmetric  task on $\Pi_2$.
Under these assumptions, we can define a weaker notion of reduction involving 
	a ``symmetrization  of $T_2$ inside  $\Pi_1$''.
	
Formally, for any subset $\Pi$ of $\Pi_1$ of cardinality $n_2$ and any one-to-one 
	mapping $\Phi$ from $\Pi_2$ onto $\Pi$, consider the task $^{\Phi}\!T_2$ for 
	$\Pi$.
Since $T_2$ is symmetric, $^{\Phi}\!T_2$ is invariant under permutation of $\Pi$,
	and so only depends on $\Pi$ (and not on the specific choice of the mapping 
	$\Phi\ :\ \Pi_2\rightarrow \Pi$).
This allows us to denote this task $^{\Pi} T_2$ instead of $^{\Phi}\!T_2$.

\begin{definition}
We say that $T_1$ $C^*\!$-reduces to $T_2$, and we note $T_1\SC T_2$, if 
	we have 
	$$T_1 \C \{ ^{\Pi} T_2\ :\ \Pi\subseteq \Pi_1\mbox{ and } |\Pi|= n_2\}.$$
\end{definition}

Actually, in the sequel we shall deal with this notion only when $T_1$ is also symmetric.
Observe that $C^*$-reduction is an interesting notion only in the case $n_1> n_2$: when 
	$n_1 = n_2$ (resp. $n_1 < n_2$), $T_1$ $C^*$-reduces to $T_2$ iff  after any renaming
	$\Pi_1\ren \Pi_2$, the task $T_1$ $C$-reduces to $T_2$  (resp., iff $T_1$ is solvable).
	
Notice that the possibility of introducing a second notion of reducibility {\em \`a la Cook}, 
	namely the $C^*$-reduction, besides $C$-reduction, relies basically on the existence 
	of  several ``partial renamings'' $\Pi_2\ren \Pi (\hookrightarrow \Pi_1)$; it is thus inherent 
	to the distributed  nature of the computations we deal with, and  has no counterpart in 
	the classical complexity theory.

From the strong transitivity property of the $C$-reduction, we  derive that $\SC$ is a transitive relation.
It is clearly reflexive.

Finally, assume that $\Pi_2\subseteq \Pi_1$.
Then it is straightforward that for $T_1$ and $T_2$ as above, if $T_1\C T_2$ then $T_1\SC T_2$.
In Section~\ref{AC}, we shall show that  except in the case $\Pi_1= \Pi_2$, the converse does
	not generally hold.
Interestingly, in Part~II, we shall exhibit classes of distributed tasks for which these two reductions
	turn out to coincide.
	
\subsection{A first example: ${\mathbf {Cons(n+f,f)}}$ is  ${\mathbf C^*\!}$-reducible to  ${\mathbf{AC(n,f)}}$}

We now see a first example of $C^*\!$-reduction, showing how to extract Consensus from 
	Atomic Commitment.
	
Let $\Pi$ be a set of $n+f$ process names.
We consider the $m =\binom{n+f}{n}$ subsets of $\Pi$ of cardinality $n$.
Let us fix an arbitrary order on these subsets $\Pi_1,\cdots,\Pi_m$, and
	a set of sanctuaries $\{1,\cdots,m\}$.
In Figure~\ref{figc*red1}, we give the code of a simple Consensus algorithm 
        for $\Pi$ using the 
	oracles ${\cal O}.\AC(\Pi_1,f), \cdots, {\cal O}.\AC(\Pi_m,f)$.  
Informally, every process $p$ consults these oracles with its initial value $v_p$, 
	according to the order $1,\cdots,m$, and skipping the indexes $i$ 
	for which $p\notin \Pi_i$.
As soon as $p$ gets a response from an oracle, it broadcasts it in $\Pi$.
Eventually, it knows all the values answered by the oracles (including those that it has 
	not consulted), and then decides on the greatest value.
	
\begin{figure*}[t]\label{figc*red1}\small 
$\hrulefill$
\begin{tabbing}
mm\=mm\=mm\=mm\=mm\=mm\=mm\=mm\=mm\=mm\=mm\= \kill

$\mathbf{Code\ for\ process\ p:}$\\ \\

\textbf{initialization:}\\
\> $d_p\in V\cup \{\bot\}$, initially $\bot$\\ \\
\textbf{for} $i=1$ \textbf{to} $m$ \textbf{do}:\\
\> \textbf{if} $p\in \Pi_i$ \textbf{then}\\
\> \> $\query ({\cal O}.\AC(\Pi_i,f)) \langle v_p \rangle $\\
\> \> $\answ ({\cal O}.\AC(\Pi_i,f)) \langle w_i \rangle $\\
\> \> $\send  \langle (i,w_i) \rangle $ to all\\
\textbf{wait until} [$\rec \langle (i, w_i) \rangle $ for all $i\in\{1,\cdots,m\}$]\\
$d_p:=\max _{i=1\cdots,m} (w_i)$\\
$\dec(d_p)$
\end{tabbing}
\vspace{-0.5cm}
\caption{A $C^*\!$-reduction from $\Cons(n+f,f)$ to $\AC(n,f)$.}
\label{reduction}
$\hrulefill$
\end{figure*}
                      
\begin{theorem}\label{c*red1}
Let $n,f$ be any positive integers, $1\leq f\leq n-1$, and let $\Pi$ be a set of
	$n+f$ processes.
The algorithm in Figure~2  solves the task $\Cons(\Pi,f)$, and so
	 $\Cons(n+f,f)$ $C^*\!$-reduces to $\AC(n,f)$.
\end{theorem}

\begin{Proof}
We first prove the termination property.
By a simple induction on $i$, we easily show that every oracle ${\cal O}.\AC(\Pi_i,f)$
	is consulted by at least $|\Pi_i|-f=n-f$ processes, and so no process is blocked in
	the sanctuary $i$.
Every correct process $p\in \Pi_i$ thus gets an answer from the oracle ${\cal O}.\AC(\Pi_i,f)$,
	and then broadcasts it in $\Pi$.
Since $n\geq f+1$, the subset $\Pi_i$ contains at least one correct process.
Therefore every correct process eventually knows the $m$ values answered by 
	the oracles ${\cal O}.\AC(\Pi_1,f), \cdots, {\cal O}.\AC(\Pi_m,f)$, and then makes a decision.
	
Irrevocability is obvious.
Agreement follows from the decision rule and the fact that every process which 
	makes a decision knows the values answered by all the oracles.  

For validity, if all the initial values are $0$, then every
	oracle is  queried with value 0 by at least one process, and so answers value 0.
Therefore, the decision value is $0$.

Suppose now that all the processes in $\Pi$ start with initial value 1. 
Since at most $f$ processes are faulty, there is at least
	one subset $\Pi_i$ in which all processes are correct.
Then the answer given by the oracle ${\cal O}.\AC(\Pi_i,f)$ is 1.
By the decision rule, it follows that the decision value is 1.\qed
\end{Proof}


\section{${\mathbf{C^{(*)}}}$-reductions when varying the number of processes}

We now investigate various $C$- and  $C^*$-reductions between tasks associated to the same
	agreement problem, but which differ in the cardinalities of the sets of processes for which they 
	are defined (and in their resiliency degrees as well).
The reductions we describe are simple;  their correctness proofs are straightforward and will be
	omitted.  
	 
The first reduction we shall give solves $\AC(n+1,f)$ using two oracles of $\AC(n,f)$ type.
Interestingly,  only one such oracle is not sufficient to solve $\AC(n+1,f)$.
In other words, we prove that although $\AC(n+1,f)$ is $C^*$-reducible to $\AC(n,f)$, it is not 
	$C$-reducible to $\AC(n,f)$.
	
We then establish that $\Cons(n,f)$ falls between $\Cons(n+1,f+1)$ and  $\Cons(n+1,f)$
	with regard to the ordering $\C$.

\subsection{$\mathbf{AC(n+1,f)}$  $\mathbf{C^*\!}$-reduces 
        but does not $\mathbf{C}$-reduce to $\mathbf{AC(n,f)}$}\label{AC}

Let $\Pi$ be any set of $n+1$ processes; each process $q\in \Pi$ 
	has an initial value $x_q\in \{0,1\}$.
Let us consider  any two different subsets  $\Pi'$ and $\Pi''$ of $\Pi$ with  $n$ processes.
(Without loss of generality, we could have assumed that $\Pi=\{1,\cdots,n+1\}$, 
	$\Pi'=\{1,\cdots,n\}$, and $\Pi''=\{2,\cdots,n+1\}$).

We first sketch a simple algorithm running on  $\Pi$ which uses both ${\cal O}.\AC(\Pi',f)$  
	and ${\cal O}.\AC(\Pi'',f)$: 
Every process $q$ first queries  ${\cal O}.\AC(\Pi',f)$ if $q\in\Pi'$, and then queries
	${\cal O}.\AC(\Pi'',f)$ if $q\in\Pi''$.
Each of these two oracles is consulted by at least $n-f$ processes, and so eventually  answers. 
Let $d'$ and $d''$ be the responses of ${\cal O}.\AC(\Pi',f)$ and ${\cal O}.\AC(\Pi'',f)$, respectively.
Every process in $\Pi'$ that is still alive sends $d'$ to all processes in $\Pi$; similarly,
	 every process in $\Pi''$ broadcasts $d''$. 
As $f<n$, $\Pi'$ and $\Pi''$ both contain at least one correct process, and so every process in $\Pi$ 
	eventually knows  both $d'$ and $d''$.
Finally, every alive process decides on $d=\min(d',d'')$.
This establishes:

\begin{proposition}\label{res2*}
If $n$ and $f$  are  two integers such that $1\leq f \leq n-1$, then $\AC(n+1,f)$ is
	$C^*$-reducible to $\AC(n,f)$.
\end{proposition}

We shall now prove that $\AC(n+1,f)$ is not solvable just using the oracle ${\cal O}.\AC(n,f)$. 
This result will demonstrate that $C$-reducibility  is actually a stronger notion than $C^*$-reducibility.

\begin{proposition}\label{nored0}
If $n$ and $f$  are  two integers such that $1\leq f \leq n-1$, then $\AC(n+1,f)$ is not 
	$C$-reducible to $\AC(n,f)$.
\end{proposition}

\begin{Proof}
As before, let $\Pi$ be a set on $n+1$ processes and $\Pi'$ a subset of $\Pi$ with
	$n$ processes.
	
For the sake of contradiction, suppose that there is an algorithm $R$ using the oracle
	${\cal O}.\AC(\Pi',f)$ which solves $\AC(\Pi,f)$.
Let $p$ be the unique process in $\Pi\setminus\Pi'$.
Consider a run $\rho=<\!\! F, I, H \!\!>$ of $R$ such that, for any $q\in\Pi$, 
	$I(q)=s_q^1$, and for any $t\in {\cal T}$, $F(t)=\{p\}$. 
In other words, $\rho$ is a run of $R$ in which all processes start with
	initial value 1 and no process is faulty except $p$ which initially crashes.
Let $d$ denote the decision value in $\rho$.

We now prove that $d=0$.
For that, we introduce  the mapping $I'$ on $\Pi$ which is identical to $I$ over 
	$\Pi\setminus \{p\}$ and satisfies $I'(p)=s_p^0$, 
	and we consider $\rho'=<\!\! F, I', H \!\!>$. 
We claim that $\rho'$ is a run of $R$.
Since $\rho$ is a run, it is straightforward that $\rho'$ satisfies R1, R2, R3, R5, and R6.
By an easy induction, we see that for any process $q$, $q\neq p$, the sequence 
	of the local states reached by $q$ are the same in $\rho'$ as in $\rho$.
This ensures that every step in $H$ is feasible from $I'$, and so R4 holds in $\rho'$.  	
Thus, $\rho'$ is a run of $R$, and by the validity condition of Atomic Commitment, 
	the only possible decision value in $\rho'$ is 0. 
This shows that $d=0$. 

Next we  construct a failure free run of $R$ for which the history begins as 
	$H$, up to the moment all processes in $\Pi'$ have made a decision.
To achieve that, we need the following lemma, where  $F_0$ denotes  the failure pattern 
	with no failure (defined formally by $F_0(t)=\emptyset$, for any $t\in {\cal T}$), and
	$H[0,t]$ denotes the prefix in $H$ of events with time less or equal to $t$.	
\begin{lemma}\label{ext1}
For any $t_0\in {\cal T}$, there exists an extension $H_0$ of $H[0,t_0]$
	such that $<\!\! F_0, I, H_0 \!\!>$ is a failure free run of $R$.
\end{lemma}
\begin{Prooflemma} 
The history $H_0$ is constructed in stages, starting from $H[0,t_0]$.
Each stage consists in adding zero or one event.
A queue of the processes in $\Pi$ is maintained, initially in an arbitrary order, and
	the messages in $\beta$ are  ordered according to the time the messages were
	sent, earliest first.
	
Suppose that the finite history $H_0[0,t]$ extending $H[0,t_0]$ is constructed.
Let $t^+$ denote the successor of $t$ in ${\cal T}$, and let $q$ be the first process 
	in the process queue.
After $H_0[0,t]$, $q$ may achieve only one type $T$ of event. 
There are three cases to consider:

\begin{enumerate}
\item $T=S$ or $T=Q$.
The automaton $R(q)$ entirely determines the event $e=(\beta, q,t^+,S,m)$ or 
	$e=(\AC(\Pi',f), q,t^+,Q,v)$ which $q$ may achieve at time $t^+$.
\item $T=R$.
In this case, the message buffer $\beta$ contains at least one message for $q$.
Then we let $e= (\beta, q,t^+,R,m)$, where $m$ denotes the earliest message for $q$ 
	in $\beta$.
\item $T=A$.
Form the successive consultations of ${\cal O}.\AC(\Pi',f)$ in $H_0[0,t]$, and focus on 
	 the latter consultation. 
There are three subcases:
	\begin{description}
	\item {\em Case 1:}  ${\cal O}.\AC(\Pi',f)$  has yet answered some value $d$.\\
	In this case, we let $e=(\AC(\Pi',f), q,t^+,A,d)$.
	\item {\em Case 2:}  ${\cal O}.\AC(\Pi',f)$ has not yet answered,
	but has been queried by all  processes~in~$\Pi'$.\\
	We let $e=(\AC(\Pi',f), q,t^+,A,d)$, where $d$  denotes the minimum of all the query values.
	\item {\em Case 3:} ${\cal O}.\AC(\Pi',f)$  has not yet answered and has not yet 
	been queried by some  processes in $\Pi'$.\\
	In this case, we skip $q$'s turn and no event is determined in this stage.
	\end{description}
\end{enumerate}

If the above procedure determines an event $e$, then we let $H_0[0,t^+] = H_0[0,t]; e$ 
        (where semicolon denotes concatenation). 
Otherwise we are in Case 3.3, and we let $H_0[0,t^+]= H_0[0,t]$.
Process $q$ is then moved to the back of the process queue.

This  inductively defines $H_0$.
By construction,   $\rho_0=<\!\! F_0, I, H_0 \!\!>$ satisfies R1--6, and so is a failure free run 
	of $R$. \nopagebreak \hfill $\Box_{Lemma 7.3}$ \nopagebreak
\end{Prooflemma}

We now instantiate  $t_0$  to be the time when the last process makes a decision in $\rho$.
The lemma provides an extension $H_0$ of $H[0,t_0]$
	such that $\rho_0=<F_0,I,H_0>$ is a run of $R$.
The decision value in $\rho_0$ is 0, which contradicts the fact that processes 
	must decide on 1 in a failure free run of an Atomic Commitment algorithm in 
	which all processes start with initial value 1.\qed
\end{Proof}

\subsection{$\mathbf{Cons(n+1,f)}$ is $\mathbf{C}$-reducible to $\mathbf{Cons(n,f)}$}

Contrary to what happens with Atomic Commitment, when dealing with Consensus, 
	a decision value for a restricted subset of processes may always be adopted 
	by all processes to make a global decision.
In other words, a process kernel may impose a common decision on the whole system 
	without violating validity for a general consensus.
	
Let $\Pi$ be any set of $n+1$ processes; each process $p\in \Pi$ 
	starts with an initial value $x_p\in \{0,1\}$.
We fix  any subset $\Pi'$ of  $\Pi$ with $n$ processes, and we consider the oracle 
	${\cal O}.\Cons(\Pi',f)$ which may be consulted by any  member of the  process kernel $\Pi'$.
A $C$-reduction  from $\Cons(\Pi,f)$ to $\Cons(\Pi',f)$ is as follows:
Every process $p$ in $\Pi'$ first queries  ${\cal O}.\Cons(\Pi',f)$ with its initial value $x_p$.
All the correct processes in $\Pi'$ eventually get a common answer $d$ since 
        at most $f$ processes 
	may be prevented from querying the oracle ${\cal O}.\Cons(\Pi',f)$.
Then every process in $\Pi'$ that is still alive sends $d$ to all processes in $\Pi$. 
As $f \leq n-1$, $\Pi'$ contains at least one correct process, and so every process in $\Pi$ 
	eventually receives $d$.
Finally, every alive process decides on $d$.
By  property  $\mathrm{O_{Cons}}$  (Section~\ref{O.T}),  validity of Consensus is satisfied. 
This establishes:

\begin{proposition}\label{res2}
If $n$ and $f$  are  two integers such that $1\leq f \leq n-1$, then $\Cons(n+1,f)$ is
	$C$-reducible to $\Cons(n,f)$.
\end{proposition}

\subsection{$\mathbf{Cons(n,f)}$ is $\mathbf{C}$-reducible to $\mathbf{Cons(n+1,f+1)}$}

At this point, one may wonder whether conversely, $\Cons(n,f)$ is $C$-reducible
	to $\Cons(n+1,f)$.
A negative answer to this question will be given in Part~II, thanks to the introduction 
	of the new class of $k$-{\em Threshold Agreement}  tasks~\cite{CL04}.
	
Instead of comparing $\Cons(n,f)$ with $\Cons(n+1,f)$, we may consider the 
	{\it a priori} harder task $\Cons(n+1,f+1)$, and show that $\Cons(n,f)$ is 
	indeed $C$-reducible to 	$\Cons(n+1,f+1)$.
	
Let $\Pi$ be any set of $n+1$ processes, and let $\Pi' $ be any subset of $ \Pi$ 
	with  $n$ processes.
The $C$-reduction from $\Cons(\Pi',f)$ to $\Cons(\Pi,f+1)$ is trivial:
Each process in $\Pi'$ just needs to query the oracle ${\cal O}.\Cons(\Pi,f+1)$ 
        with its initial value.
The oracle definitely answers since it is consulted by at least $n-f=(n+1)-(f+1)$ processes.
Every process finally decides on the value provided by ${\cal O}.\Cons(\Pi,f+1)$.
By property $\mathrm{O_{Cons}}$,  validity of Consensus is satisfied. 
This establishes:
	
\begin{proposition}\label{res1}
If $n$ and $f$  are  two integers such that $1\leq f  \leq  n-1$, then $\Cons(n,f)$ is
	$C$-reducible to $\Cons(n+1,f+1)$.
\end{proposition}

Proposition~\ref{res1} states that $\Cons(n+1,f+1)$ is at least as 
        hard as $\Cons(n,f)$, which, by Proposition~\ref{res2}, is 
	at least as hard as $\Cons(n+1,f)$.
In other words, $\Cons(n,f)$ is sandwiched between $\Cons(n+1,f)$ and $\Cons(n+1,f+1)$
	with respect to the ordering $\C$.
	
By Proposition~\ref{res1}, it follows that if $\Cons(n,n-1)$ is $C$-reducible to some task $T$, then
	every task $\Cons(m,m-1)$,  with $m$ not greater than $n$, also $C$-reduces to $T$.
Thus to any distributed task $T$,  it is natural to associate  the largest positive integer $n$ such that 
	$\Cons(n,n-1)\C T$.
Pursuing the analogy between oracles and shared objects that we have outlined in 
	Section~\ref{relnot}, this number actually corresponds to 
	the {\em consensus number} defined by Herlihy in~\cite{Her91}.
 
\section{${\mathbf C}$-reduction between Atomic Commitment and Consensus}

This section is devoted to the $C$-reducibility between Consensus and Atomic
	Commitment tasks for some given set of processes.
Our main results are impossibility results: in Sections~\ref{ac1} and~\ref{cons2},
	we show  that, for values of resiliency degree greater than one, 
	Consensus and Atomic Commitment tasks are indeed not $C$-comparable.\footnote{In 
	an unpublished joint work with S. Toueg~\cite{CT00}, a weaker version of these results
	involving only an informal notion of reduction was already obtained.
	It stated that $\AC(n,f)$ is not reducible to $\Cons(n,f)$ when $1\leq f\leq n-1$, 
	and that $\Cons(n,f)$ is not reducible to $\AC(n,f)$ when $2\leq f\leq n-1$.
	Due to the lack of a formal model for oracles, the proofs had to be of a different nature,
	and indeed were based on arguments {\em \`a la } Fischer-Lynch-Paterson~\cite{FLP85}.}
In Section~\ref{cons1}, we also prove a $C$-reducibility result from $\Cons(n,1)$  to 					$\AC(n,1)$, concerning the remaining case of resiliency degree one.
	
\subsection{Atomic Commitment cannot be reduced to Consensus}\label{ac1}

To make our result as strong as possible, we are going to prove it with a resiliency degree of the
	Atomic Commitment task as small as possible, and a resiliency degree of the Consensus 
	task as great as possible.
Actually, we prove the following theorem:

\begin{theorem}\label{noredac}
For any integer $n$, $n\geq 2$,  $\AC(n,1)$ is not
	$C$-reducible to $\Cons(n,n-1)$, and thus, for any integer $f$  such that  
	$1\leq f\leq n-1$, $\AC(n,f)$ is not
	$C$-reducible to $\Cons(n,f)$.
\end{theorem}

\begin{Proof}
Let $\Pi$ be a set of $n$ process names. 
Suppose, for the sake of contradiction, that there is an algorithm $R$ for the task $\AC(\Pi,1)$ 
	which uses the oracle ${\cal O}.\Cons(\Pi,n-1)$. 
Let $p$ be any process in $\Pi$.
Consider a run $\rho=<\!\! F, I, H \!\!>$ of $R$ such that, for any $q\in\Pi$,
      $I(q)=s_q^1$, and for any $t\in {\cal T}$, $F(t)=\{p\}$. 
In other words, $\rho$ is a run of $R$ in which all processes start with	
      initial value 1 and no process is faulty except $p$ which initially crashes.
Let $d$ denote the decision value in $\rho$.

We are going to prove that $d=0$.
For that, we introduce  the mapping $I'$ which is identical to $I$ over 
$\Pi\setminus \{p\}$ and
	satisfies $I'(p)=s_p^0$, 
	and we consider $\rho'=<\!\! F, I', H \!\!>$. 
We claim that $\rho'$ is a run of $R$.
Since $\rho$ is a run, it is straightforward that $\rho'$ satisfies R1, R2, R3, R5, and R6.
By an easy induction, we see that for any process $q$, $q\neq p$, the sequence 
	of the local states reached by $q$ are the same in $\rho'$ as in $\rho$.
This ensures that every step in $H$ is feasible from $I'$, and so R4 holds in $\rho'$. 
Thus, $\rho'$ is a run of $R$, and by the validity condition of Atomic Commitment, 
	the only possible decision value in $\rho'$ is 0. 
This shows that $d=0$. 

Now from $\rho$, we are going to construct a failure free run of $R$ by using the asynchronous 
	structure of computations.
To achieve that, we need the following lemma, where  $F_0$ denotes  the failure pattern 
	with no failure (defined formally by $F_0(t)=\emptyset$, for any $t\in {\cal T}$), and
	$H[0,t]$ denotes the prefix in $H$ of events with time less or equal to $t$.
		
\begin{lemma}
For any $t_0\in {\cal T}$, there exists an extension $H_0$ of $H[0,t_0]$
	such that $<\!\! F_0, I, H_0 \!\!>$ is a failure free run of $R$.
\end{lemma}
\begin{Prooflemma} 
The proof technique is similar to the one of Lemma~\ref{ext1}, except for Case~3 
	($T=A$).
In this case, we also form the successive consultations of ${\cal O}.\Cons(\Pi,n-1)$ in $H_0[0,t]$, 
	and focus on  the latter consultation. 
Note that  process $q$ has necessarily queried ${\cal O}.\Cons(\Pi,n-1)$
	during  this consultation; let $v$ be the value of this query.
There are two subcases:
	\begin{description}
	\item {\em Case 1:}  ${\cal O}.\Cons(\Pi,n-1)$  has yet answered some value $d$.\\
	In this case, we let $e=(\Cons(\Pi,n-1), q,t^+,A,d)$.
	\item {\em Case 2:}  ${\cal O}.\Cons(\Pi,n-1)$ has not yet answered.\\
	We let $e=(\Cons(\Pi,n-1), q,t^+,A,v)$.
	\end{description}

We complete the proof of this lemma as the one of 
	Lemma~\ref{ext1}.\nopagebreak \hfill $\Box_{Lemma 8.2}$ \nopagebreak
\end{Prooflemma}

We now instantiate  $t_0$  to be the time when the last process makes a decision in $\rho$.
The lemma provides an extension $H_0$ of $H[0,t_0]$
	such that $\rho_0=<F_0,I,H_0>$ is a run of $R$.
The decision value in $\rho_0$ is 0, which contradicts the fact that processes 
	must decide on 1 in a failure free run of an Atomic Commitment algorithm in 
	which all processes start with initial value 1.\qed
\end{Proof}

\subsection{Consensus cannot be generally reduced to Atomic Commitment}\label{cons2}

Conversely, we now prove that Consensus is generally not $C$-reducible to Atomic Commitment.
The proof technique is new and quite different from the one of Theorem~\ref{noredac}: basically,
	it consists in a ``meta-reduction'' to the impossibility result of Consensus  with one
	failure~\cite{FLP85}.
		 
As for Theorem~\ref{noredac},  to make our result as strong as possible, we state it with a resiliency 
	degree of the Consensus task as small as possible and a resiliency degree of the Atomic 
	Commitment task as great as possible. 
	
\begin{theorem}\label{noreduce}
For any integer $n$, $n\geq 3$,   $\Cons(n,2)$ is not $C$-reducible
	to $\AC(n,n-1)$, and thus for any integer $f$ such  that $2\leq f \leq n-1$, 
	$\Cons(n,f)$ is not $C$-reducible to $\AC(n,f)$.
\end{theorem}

\begin{Proof}
We also proceed by contradiction: let $\Pi$ be a set of $n$ process names, and suppose that 
	there is an algorithm $R$ for $\Cons(\Pi,2)$ using  the oracle ${\cal O}.\AC(\Pi,n-1)$.
Let $\sigma$ denote the sanctuary of this oracle.
We fix some process $p\in \Pi$.
From $R$, we shall design an algorithm $A$ running on the system 
	$\Pi\setminus\{p\}$, which uses no oracle.
We then shall prove that $A$ solves the task $\Cons(\Pi\setminus\{p\},1)$, which contradicts the 
	impossibility of Consensus with one failure established by Fischer, Lynch, and 
	Paterson~\cite{FLP85}.  
	
For each process $q$, we define the automata $A(q)$ in the following way:  	
	\begin{itemize}
	\item the set of states of $A(q)$ is the same as the one of $R(q)$;
	\item the set of initial states of $A(q)$ is the same as the one of $R(q)$;
	\item each transition $(s_q,[q,m,\bot], s'_q)$ of $R(q)$ in which $q$ consults no oracle
	 is also a transition of $A(q)$;
	\item each transition $(s_q,[q,m,1], s'_q)$ of $R(q)$ in which the oracle 
	answers 1 is removed;
	\item each transition $(s_q,[q,m,0], s'_q)$ of $R(q)$ in which the oracle 
	answers 0 is replaced by the transition $(s_q,[q,m,\bot], s'_q)$.
	\end{itemize}
Note that all the steps in  $A(q)$ are of the form  $[q,m,\bot]$; in other words, the algorithm 
	$A$ uses no oracle.

Let $\rho_A=<\!\! F, I, H \!\!>$ be any run of $A$.
Each event  in $H$  is of the form $e=(\beta,q,-,-,-)$, and is part of  some  transition 
	$(s_q, [q,m,\bot] ,s'_q)$ of $A(q)$, where $m\in M\cup \{ \mbox{null} \}$.
In the construction of $A(q)$ described above, this transition results from some unique transition 
	of $R(q)$, of the form $(s_q, [q,m,\bot] ,s'_q)$ or $(s_q, [q,m,0],s'_q)$.
In this way, to each event in $H$, we associate a unique transition of $R(q)$ in which the oracle
	is not consulted or  answers 0.
	
Now, to each run $\rho_A=<\!\! F, I, H \!\!>$ of $A$, we associate the triple  
	$\rho_R=<\!\! F', I', H' \!\!>$, where the failure pattern $F'$ is defined by 
	$$F' :  t\in {\cal T}\rightarrow F'(t)=F(t)\cup\{p\},$$
	the mapping $I'$ by: 
	\begin{enumerate}
	\item if  $I(q)=s^0_q$ for some process $q\neq p$, then $I'(p)=s^0_p$;
	otherwise $I'(p)=s^1_p$,
	\item for any process $q\in \Pi\setminus \{p\}$, $I'(q)=I(q)$;	
	\end{enumerate}
and the sequence $H'$ is constructed from $H$ by the following rules:
	\begin{enumerate}
	\item any event in $H$ that is associated to a transition of $R$ in which 
	the oracle is not consulted is left unchanged;
	\item an event $(\beta,q,t,\mbox{R},m)$ in $H$, which is associated to some transition 
	in $R(q)$ of the form  $(s_q, [q,m,0], s'_q)$, is replaced in $H'$ by  the two events series 	
	$\langle (\beta,q,t,\mbox{R},m),(\sigma,q,t,\mbox{Q},v)\rangle$,  where $v$ is the query 
	value in $s_q$; 
	\item similarly, an event $(\beta,q,t,\mbox{S},m)$ in $H$ which is associated to some 
	transition in $R(q)$ of the form  $(s_q, [q,-,0], s'_q)$, is replaced  in $H'$ by 
	$\langle (\sigma,q,t,\mbox{A},0),(\beta,q,t,\mbox{S},m)\rangle$.
	 \end{enumerate}

We claim that $\rho_R$ is a run of $R$.
By construction, there is no event in $H'$ whose process name is $p$,
	and each event in $H'$ at time $t$ corresponds to at least one event in $H$
	that also occurs at time $t$.
Since $H$ is compatible with $F$ and $F'(t)=F(t)\cup \{p\}$, it follows that $H'$
	is compatible with $F'$.
For any process $q\in \Pi$,  $H|q$ is well-formed, and so  is $H'|q$.
This proves that $H'$ satisfies R2.

From the R3, R4, and R6 conditions for $H$, it is also  immediate to prove that in turn, 
	$H'$ satisfies  R3, R4, and R6.
	
Now since $F(t)\subseteq F'(t)$, every process $q$ which is correct in $F'$ is also correct
	in $F$, and so takes an infinite number of steps in $H$.
By construction of $H'$, it follows that $q$ takes an infinite number of steps in $H'$.
Thus $H'$ satisfies R5.

Finally, to show that $\rho_R$ satisfies R1, we  focus on a consultation of $\sigma$ in $H'$. 
By construction, the only possible value answered by the oracle is 0.
This trivially enforces agreement.
For validity of atomic commitment, since there is a faulty process in $F'$, 
	the answer 0 is allowed for $F'$ and any input vector $\vec{V}\in \{0,1\}^{\Pi}$.
Every step in $H$ is complete (with a receipt and a state change), and so 
	by construction of $H'$, the oracle answers to each query in $H'$. 
It follows that $H'|\sigma$ is an history of the oracle ${\cal O}.\AC(\Pi,n-1)$.
This completes the proof that $\rho_R=<\!\! F', I', H' \!\!>$  is a run of $R$.

Let $\rho_A$ be any run of $A$ with at most one failure;
	 in the corresponding run $\rho_R$ of $R$,  at most two processes fail .
Since $R$ is an algorithm for $\Cons(\Pi,2)$,
	$\rho_R$ satisfies the termination, agreement, irrevocability and validity 
	conditions of Consensus.
It immediately follows that the run $\rho_A$ which $\rho_R$ stems from also satisfies 
	the termination, agreement, and irrevocability conditions.
Moreover, by definition of $I'$, if all processes start with the same initial value 
	$v$ in $\rho_A$, then they also have the same initial value $v$ in $\rho_R$; 
	the only possible decision value in $\rho_R$, and so in $\rho_A$, is $v$.
	
Consequently,  $A$ is an algorithm for $\Cons(\Pi\setminus\{ p\},1)$  using no oracle, a contradiction 
 	with~\cite{FLP85}.\qed
 \end{Proof}

\subsection{Resiliency degree 1}\label{cons1}

Theorem~\ref{noreduce} establishes that $\Cons(n,f)$ is not $C$-reducible
	to $\AC(n,f)$ when $f>1$.
In this section, we go over the remaining case $f=1$: we show that 
	if $n>2$, then $\Cons(n,1)$ is $C$-reducible to $\AC(n,1)$.

In Figure~\ref{figf=1}, we give an algorithm using an Atomic Commitment 
	oracle which solves Consensus in a system $\Pi$ with $n>2$ processes if 
	at most one crash occurs.
Our Consensus algorithm uses the Atomic Commitment oracle only once and
	only to get some informations about failures.
More precisely, with the help of this oracle, processes 
	determine whether some failure has occurred before each process
	sends its initial value for Consensus.
If no failure is detected, then every process waits until it receives the initial value
	from every process, and Consensus is easily achieved in this case.
Otherwise the oracle indicates an eventual failure.
The oracle  ${\cal O}.AC(\Pi,1)$ allows us to make accurate failure detection (that is no false
	detection), but the delicate point lies in the fact that the failure may occur at any time, 
	possibly in the future.
We thus had to devise the second part of the algorithm in order to deal with this lack of information
	about the time when the failure occurs.
For that, we have ``de-randomized'' Ben-Or algorithm~\cite{Ben83}: instead of tossing a coin at some
	points  of the computation, processes adopt the fixed value 0.
In the resulting algorithm, the occurrence of the failure enforces correct processes to make a decision.
The complete code of the $C$-reduction is given in Figure~\ref{figf=1}.

\begin{figure*}[t]\small 
$\hrulefill$
\begin{tabbing}
mm\=mm\=mm\=mm\=mm\=mm\=mm\=mm\=mm\=mm\=mm\= \kill
\textbf{Variables of process} $p:$\\ \\
\> $x_p\in V$, initially $v_p$\\
\> $r_p\in \N$, initially 1\\ \\
\textbf{Algorithm for  process} $p:$\\ \\
\> $\send  \langle v_p \rangle $ to all\\
\> $\query ({\cal O}.\AC(\Pi,1)) \langle 1 \rangle $\\
\> $\answ({\cal O}.\AC(\Pi,1)) \langle d\rangle $\\
\>\textbf{if} $d=1$\\
\>\textbf{then}\\
\>\> \textbf{wait until} [$\rec \langle v_q \rangle $ from all $q\in\Pi$]\\
\>\> $x_p:=\min _{q\in\Pi} (v_q)$\\
\>\> $\dec(x_p)$\\
\>\textbf{else}\\
\>\> \textbf{repeat} \\
\>\> \> $\send  \langle (R,x_p,r_p) \rangle $ to all\\
\>\>\> \textbf{wait until} [$\rec \langle (R,*,r_p) \rangle $ from $n-1$ processes] 
                                      (where $*$ can be 0 or 1)\\
\>\>\> \textbf{if} more than $n/2$ messages have the same value $v\in \{0,1\}$ 
          in the second component\\
\>\>\> \textbf{then}\\
\>\>\>\> $\send  \langle (P,v,r_p) \rangle $ to all\\
\>\>\> \textbf{else}\\
\>\>\>\> $\send  \langle (P,?,r_p) \rangle $ to all\\
\>\>\> \textbf{wait until} [$\rec \langle (P,*,r_p) \rangle $ from $n-1$ processes] 
                                      (where $*$ can be 0, 1, or ?)\\
\>\>\> \textbf{if} at least two of the $\langle (P,*,r_p) \rangle $'s received have the same $w\in\{0,1\}$
         in the second component \\
\>\>\> \textbf{then}\\
\>\>\>\>  $x_p:=w$\\
\>\>\>\>  $\dec(w)$\\
\>\>\> \textbf{else}\\
\>\>\> \>\textbf{if}  one of the $\langle (P,*,r_p) \rangle $'s received have $w\in\{0,1\}$
         in the second component \\ 
\>\>\>\>\textbf{then}\\
\>\>\>\>\> $x_p:=w$\\
\>\>\>\> \textbf{else}\\
\>\>\>\>\> $x_p:=0$\\
\>\>\> $r_p:=r_p +1$\\
\end{tabbing}
\caption{A $C$-reduction from  $\Cons(n,1)$ to $\AC(n,1)$}
\label{figf=1}
$\hrulefill$
\end{figure*}

\begin{theorem}\label{t=1}
For any integer $n>2$, $\Cons(n,1)$ is $C$-reducible to $\AC(n,1)$.
\end{theorem}

\begin{Proof}
Let  $\rho=<\!\! F, I, H \!\!>$ denote a run of the algorithm in Figure~\ref{figf=1}.
To prove that $\rho$ satisfies the termination, irrevocability, agreement, and 
	validity conditions of Consensus, we shall distinguish the case in which 
	the oracle ${\cal O}.\AC(n,1)$ answers 0 from the one in which it answers 1 ($d=0$ 
	and $d=1$).\\

\noindent\textit{Case} $d=1$.  
Irrevocability and validity are obvious.

For termination and agreement, we claim that  in this case, all processes query 
	the oracle ${\cal O}.AC(\Pi,1)$ in $\rho$.
In proof, if some process $p$ does not consult  the oracle, then 
	 by  property $\mathrm{O_{AC}}$ (cf. Section~\ref{O.T}),  the oracle ${\cal O}.AC(\Pi,1)$ must
	 answer $0$, a contradiction.
	
Since before consulting the oracle, every process has to broadcast its initial value,
	all processes that are still alive do receive the $n$ initial values for Consensus.
Termination and agreement conditions easily follow.\\

\noindent\textit{Case} $d=0$.
First, we claim that $\rho$ satisfies the validity condition of Consensus. 
In proof, suppose that all processes start with the same initial value $v$.
Every process sends $(R,v,1)$ to all;  since $n>2$, every process
	proposes value $v$ at the first round, i.e., sends $(P,v,1)$ to all.
As $n>2$, it follows from the code that each process then decides  $v$.
 
For agreement, we argue as for Ben-Or algorithm.
First, because of the majority rule which determines when a process proposes value $v\in\{0,1\}$
	(i.e., sends $(P,v,r)$ to all), it is impossible for a process  to propose 0 and for another one to
	propose 1 in the same round.
Suppose that some process makes a decision in $\rho$, and let $r$ denote the first round
	at which a process decides.
If process $p$ decides $v$ at round $r$, then it has received at least 2 propositions for
	$v$ at round $r$.
Thus, every process $q$ receives at least one proposition for $v$ at round $r$,
	and so we have $x_q=v$ at the end of round $r$.
This enforces every process to decide  $v$ at the latest at round $r+1$, and to keep
	deciding $v$ in all subsequent rounds.
In other words, $\rho$ satisfies agreement and irrevocability.

We now argue termination.
Since every query value of ${\cal O}.AC(\Pi,1)$ is 1,  this implies that 
	exactly one failure occurs in run $\rho$.
For every process $p$, we consider  the round  $r_p$  process $p$ is executing when this 
	failure occurs, and we let $r_{\rho}=\max_{p\in Correct(F)}(r_p) + 1$.
Suppose no process has made a decision by the end of round $r_{\rho}$.
All correct processes receive the same set of $n-1$ messages of the form $(R,-,r_{\rho})$,
	and so they propose the same value $v\in\{0,1,?\}$ at round $r_{\rho}$.
If $v\neq ?$, then every correct process decides $v$ since it receives $n-1\geq 2$ propositions
	for $v$.
Otherwise, $v=?$ and every correct process $p$ sets $x_p$ to 0.
Since $n>2$, it is easy to see that  in this case, correct processes decides 0 at round $r_{\rho} +1$.
This completes the proof of termination.\qed
\end{Proof}

Note that the reduction above is much stronger than the one given by Theorem~\ref{c*red1}
	for the particular case $f=1$.
Firstly, Theorem~\ref{t=1} establishes that $\AC(n,1)$ is harder to solve than $\Cons(n,1)$
	whereas Theorem~\ref{c*red1} just compares $\AC(n,1)$ with $\Cons(n+1,1)$, which is
	shown to be a weaker task than $\Cons(n,1)$ by Proposition~\ref{res2}.
Secondly, Theorem~\ref{t=1} is a $C$-reduction result, and not only a $C^*$-reduction result
	as Theorem~\ref{c*red1} is.

We end this section by deriving an interesting corollary from 
	Theorems~\ref{nok-TAgred}~and~\ref{t=1}.
On one hand, Theorem~\ref{nok-TAgred} shows that $\Cons(n,1)$ is not $K$-reducible to
	$\AC(n,1)$.
On the other hand, Theorem~\ref{t=1} establishes that $\Cons(n,1)$  $C$-reduces to $\AC(n,1)$.
Hence, there are two {\em unsolvable} tasks for which the reductions $\K$ and $\C$ differ.
In other words, we have proved that {\em in distributed computing, $K$-reduction is strictly 
	stronger than  $C$-reduction} (compare with~\cite{LLS75}, where the relations between various 
	polynomial-time reducibilities in classical complexity theory are examined).
	
\subsection{Extracting Consensus from Atomic Commitment and vice-versa}

Theorem~\ref{noreduce} shows that an oracle for $\AC(n,f)$ does not help to solve $\Cons(n,f)$,
	and more generally to solve any Consensus task for $\{1,\cdots,n\}$.
However, Theorem~\ref{c*red1} partially gets around this impossibility result by enlarging the set 
	of processes, and so by weakening the Consensus task to be solved 
	(cf. Proposition~\ref{res2}).
Indeed, this theorem asserts  that if we grant the processes in $\{1,\cdots, n+f\}$ the 
	ability to query oracles of type ${\cal O}.\AC(n,f)$, then $f$-resilient Consensus is a solvable
	task.\footnote{Thanks to the introduction of the $k$-Threshold Agreement tasks, we shall prove 
	a better result in Part~II, namely that $\Cons(n+f-1,f)$ $C^*$-reduces to $\AC(n,f)$.}
In other words, contrary to $\Cons(n,f)$, the task $\Cons(n+f,f)$ can be extracted from 
        $\AC(n,f)$.

Conversely, we may wonder whether enlarging the set of processes $\{1,\cdots,n\}$ could make
	Atomic Commitment tasks solvable if we grant the processes to consult  oracles of type
	${\cal O}.\Cons(n,f)$.
In fact, as an application of our previous results, we may prove that no $f$-resilient Atomic Commitment 		task can be extracted from ${\cal O}.\Cons(n,f)$:
\begin{proposition}
For any integers $n$, $m$, and $f$ such that $1\leq f\leq n-1$ and $n\leq m$, $\AC(m,f)$ is not 
	$C^*$-reducible to $\Cons(n,f)$.
\end{proposition}
\begin{Proof}
For the sake of contradiction, assume that $\AC(m,f)\SC\Cons(n,f)$, that is 
	$$\AC(m,f)\C \{\Cons(\Pi,f)\ :\ \Pi\subseteq \{1,\cdots,m\}\mbox{ and } |\Pi|=n\}.$$
By Proposition~\ref{res1} applied $m-n$ times, each $\Cons(\Pi,f)$ task  $C$-reduces to 
	$\Cons(m,f+m-n)$.
Using strong transitivity of the $C$-reduction, we get that $\AC(m,f)\C\Cons(m,f+m-n)$.
As $n\leq m$, we trivially have $\Cons(m,f+m-n)\C\Cons(m,f)$, and so it follows that 
	$\AC(m,f)\C\Cons(m,f)$,
	which contradicts  Theorem~\ref{noredac}.\qed
\end{Proof}

Roughly speaking, this proposition states that Consensus contains no Atomic Commitment component.
Together with the reducibility result in Theorem~\ref{c*red1} alluded above, this corroborates the
 	popular belief that Consensus is easier to solve than Atomic Commitment.

\subsection*{Acknowledgments}

It is a pleasure to thank Andr\'e Schiper and Gerard Tel  for helpful questions and advice during 
	the writing of this paper.
I am also grateful to Sam Toueg  for valuable discussions on Consensus and Atomic Commitment
	 problems, and  to Gadi Taubenfeld and Shlomo Moran for communicating their papers.
	 

\end{document}